\documentclass[11pt]{amsart}
\usepackage{geometry}                
\usepackage{graphicx}
\usepackage{amsmath,amssymb}
\usepackage{epstopdf}
\usepackage{subfig}
\usepackage{xcolor}
\usepackage[normalem]{ulem}

\usepackage{natbib}

\newcommand{\vect}[1]{\boldsymbol{#1}}

\newcommand{\curl}[1]{\boldsymbol{\nabla}\times #1}

\title{Inertial modes in Near-spherical geometries}
\author{J. Rekier, A. Trinh, S. A. Triana, V. Dehant}
\date{October 29, 2018}

\begin{document}
\maketitle
\begin{abstract}
We propose a numerical method to compute the inertial modes of a container with near-spherical geometry based on the fully spectral discretisation of the angular and radial directions using spherical harmonics and Gegenbauer polynomial expansion respectively. This allows to solve simultaneously the Poincar\'e equation and the no penetration condition as an algebraic polynomial eigenvalue problem. The inertial modes of an exact oblate spheroid are recovered to machine precision using an appropriate set of spheroidal coordinates. We show how other boundaries that deviate slightly from a sphere can be accommodated for with the technique of \emph{equivalent spherical boundary} and we demonstrate the convergence properties of this approach for the triaxial ellipsoid.\\
{{\it\bf keywords :} Planetary interiors; Core; Numerical solutions.}
\end{abstract}

\section{Introduction}

Coriolis forces in a rotating fluid support oscillatory motions known as inertial waves. When the fluid is rotating inside a boundary, as is the case for many astrophysical objects including our planet, inertial \emph{modes} can exist. Experimentally, these modes can be excited mechanically by means of libration of the bounding surface \citep{aldridge1969}, by precession \citep{malkus1968}, by tidal forces \citep{morize2010}, or by the differential rotation of a solid inner core \citep{kelley2007}. Although the first theoretical studies of inertial modes date back towards the end of the 19th century, with the works of \citet{thomson1880, poincare1885} and \citet{bryan1889}, the role of inertial modes in the dynamics of rotating stars, planets and other astrophysical bodies remains largely unexplored.  

With an eye towards the dynamics of bodies deformed by rotation and tides, it is relevant to obtain the solutions for inertial mode oscillations of rotating fluids within \emph{near-spherical} boundaries such as spheroids or triaxial ellipsoids. In the analytical realm, implicit solutions for inviscid inertial eigenmodes in a spheroid were first obtained by \citet{bryan1889} using bi-spheroidal coordinates. In a triaxial ellipsoid the first relevant work is that of \citet{hough1895}, who employed Lam{\'e} functions. Explicit solutions in the spheroid for few low degree modes were obtained much later by \citet{Kudlick1966}. Explicit solutions for all inertial eigenmodes in a spheroid were not available until the relatively recent work of \citet{zhang2004}, who also included the first order viscous corrections. In the case of a triaxial ellipsoid, \citet{vantieghem2014} provided an algorithm to construct analytical inviscid solutions that are linear and quadratic in the cartesian coordinates, with the possibility to include higher order solutions. The long standing question about the completeness of the inertial modes to represent any smooth fluid flow within a full rotating ellipsoid was answered affirmatively by the work of \citet{backus2017} and \citet{ivers2017a}. It turns out that inertial modes are also complete when the rotation axis is arbitrary with respect to the principal axes of the ellipsoid \citep{ivers2017b}.

In the numerical realm, studies of inertial eigenmodes in ellipsoids are based predominantly on finite element methods \citep[e.g.][]{chan2010,cebron2010b, vantieghem2014}. The works of \citet{schmitt2006} and \citet{schmitt2004} are  perhaps the only ones using a set of oblate spheroidal coordinates. All these numerical studies usually focus on the time evolution of the flow field inside the spinning ellipsoid, which is in turn subject to some kind of additional mechanical forcing (e.g. libration) to excite inertial modes.


The purpose of this paper is the numerical computation of the inertial modes of a fluid rotating inside an ellipsoidal boundary numerically using a fully spectral discretisation. The radial direction is discretised using a relatively recent technique which employs Chebyshev and Gegenbauer polynomials \citep{olver2013}. This has the advantage to result in a sparse matrix representation of differential operators. The discretisation in the angular directions is carried out using a spherical harmonics decomposition. 
For the special cases of an exactly spherical or oblate spheroidal container, the spectral decomposition can be used to recover the inertial modes with an accuracy that is limited only by numerical precision. In order to treat the triaxial case, we extend the equivalent spherical domain technique introduced by \citet{smith1974} and substantiate it with an analysis of numerical convergence. This makes it possible, at least in principle, to compute the inertial modes for any boundary that deviates only slightly from sphericity. 
We also demonstrate a technique to recover semi-analytical solutions in an oblate spheroid and in a triaxial ellipsoid using systems of bi-spheroidal and bi-ellipsoidal coordinates. This allows us to cross-validate both approaches.

Our numerical method paves the way for future extensions of the model to accommodate features that the analytical solutions are not capable of, such as the inclusion of viscosity, magnetic effects or stratification and, most prominently, the effect of boundary topography.

The inviscid problem that we present in this study is an interesting instance of a \emph{polynomial eigenvalue problem} with a non-standard boundary condition. This problem also serves as a concrete example demonstrating the potential of a more general method that we have developed to treat similar problems consisting of general systems of \emph{Partial Differential Equations} (PDEs) in \emph{near-spherical geometries}. 

The paper is structured as follows. In Sec.~\ref{sec:method}, we review the mathematical description of inertial waves inside a cavity before we explain the methods used to solve the resulting equations numerically. In Sec.~\ref{sec:results}, we demonstrate the validity of our approach by treating the case of a flow within an ellipsoidal boundary. Sec.~\ref{sec:discussion} provides a discussion of how the method can easily be extended to more complex systems. The Appendix at the end of this paper provides details on how to compute the analytical solutions to the inertial modes as well as other useful technical information used throughout the present paper.

\section{Method}
\label{sec:method}

\subsection{Mathematical description of inertial modes}

We consider an incompressible, homogeneous and inviscid fluid contained in a rigid cavity which rotates with angular velocity, $\Omega$ around the z-axis, ${\bf \hat{z}}$. In the frame attached rigidly to the rotating cavity and using $1/\Omega$ as the unit of time, the momentum balance reads
\begin{equation}
\partial_t{\bf u}+2{\bf \hat{z}}\times{\bf u}=-{\bf \nabla}p,
\label{eq:navstok}
\end{equation}
where $\bf u$ represents the flow velocity and $p$ the reduced pressure (which includes the centrifugal and gravitational potential). We are assuming that the fluid velocity is small compared to the maximum tangential velocity of the cavity so that we can discard the non-linear $({\bf u}\cdot \nabla){\bf u}$ term. We further assume that the fluid motion is periodic in time:
\begin{align}
{\bf u} &= {\bf u}_0({\bf r}){\rm e}^{2{\rm i}\lambda t}+{\bf u}^*_0({\bf r}){\rm e}^{-2{\rm i}\lambda t}~,\\
p &= p_0({\bf r}){\rm e}^{2{\rm i}\lambda t}+p^*_0({\bf r}){\rm e}^{-2{\rm i}\lambda t}~,
\end{align}
where $*$ denotes the complex conjugate. After some algebraic manipulation, Eq.~(\ref{eq:navstok}) reduces to the so-called Poincar\'e equation for the pressure amplitude $p_0$ \citep{poincare1885}: 
\begin{equation}
-\lambda^2\nabla^2 p_0+({\bf \hat{z}}\cdot{\bf \nabla})^2 p_0=0~,
\label{eq:Poincare}
\end{equation} 
where $\lambda$ is the half-frequency of the motion. We employ a no-penetration boundary condition for the velocity ${\bf u}\cdot{\bf \hat{n}}|_{\partial \mathcal{V}}=0$ at the bounding surface $\partial\mathcal{V}$, which translates into the following condition for the pressure field $p_0$ \citep{greenspan1968}:
\begin{equation}
-\lambda^2{\bf \hat{n}}\cdot{\bf \nabla}p_0+i\lambda({\bf \hat{z}}\times{\bf \hat{n}})\cdot{\bf \nabla}p_0+({\bf\hat{n}}\cdot{\bf \hat{z}})({\bf \hat{z}}\cdot{\bf\nabla}p_0)|_{\partial \mathcal{V}}=0~.
\label{eq:bcPoincare}
\end{equation}
The velocity field can be recovered from the solution for $p$ using the following expression:
\begin{equation}
{\bf u}_0\equiv\frac{1}{1-\lambda^2}\frac{1}{2}\left({\bf \hat{z}}\times{\bf \nabla}p_0-i\lambda{\bf \nabla}p_0 +\frac{i}{\lambda}({\bf \hat{z}}\cdot{\bf\nabla}p_0){\bf \hat{z}}\right)~.
\label{eq:ufromp}
\end{equation}

Finding analytical solutions to the problem just described proceeds in two steps. The first one is a change of coordinates that reduces Eq.~(\ref{eq:Poincare}) to a Laplace equation. The Laplace operator and the boundary condition must both be separable in these coordinates, something that is only possible in a limited number of coordinates. The second step is to inject the formal solution of the Laplace equation into the boundary condition and solve for $\lambda$. 
We give a detailed illustration for the cases of the sphere, the oblate spheroid and the triaxial ellipsoid in Appendix~\ref{sec:analytical}. The solutions obtained will serve us to validate the numerical results of Sec.~\ref{sec:results}.

The system of Eqs.~(\ref{eq:Poincare}-\ref{eq:bcPoincare}) can be cast numerically as an eigenvalue problem that is \emph{polynomial in the eigenvalue} $\lambda$. There are methods to solve algebraic systems of this type. In order to use these, we first turn the above differential problem into an algebraic problem of the form
\begin{equation}
(A_0\lambda^0+A_1\lambda^1+A_2\lambda^2){\bf x}=0~.
\label{eq:PEP}
\end{equation}
where the $A_i$'s are square matrices, Fig.~\ref{fig:matricess} gives a visual representation of these matrices. Sec.~\ref{sec:spectraldiscretisation} explains how these can be obtained from Eqs.~(\ref{eq:Poincare}) and (\ref{eq:bcPoincare}).

The direct numerical resolution of the Poincar\'e equation for the motion of a rotating fluid is not very common practice. The reason being that this formulation is not valid when viscosity is taken into account. When this is the case, one uses an approach based on the following decomposition of the (solenoidal) velocity field, $\vect{u}$~:
\begin{equation}
\vect{u}=\curl{\curl{(P\vect{r})}}+\curl{(T\vect{r})}~,
\end{equation}
where $P(\vect{r})$ and $T(\vect{r})$ are the \emph{poloidal} and \emph{toroidal} scalar fields respectively. After inserting the above in the momentum Eq.~(\ref{eq:navstok}), one can write two independent scalar equations for $P$ and $T$ by separately considering the radial projections of the curl of the momentum equation and the curl of its curl \citep{rieutord1997}. This latter approach can be used as well when dealing with the inviscid problem so long as the domain of integration encloses the origin of coordinates, \emph{i.e.}, containers that have the topology of the spherical shell are prohibited \citep{rieutord2000}.

Both methods have their own advantages and drawbacks. The formulation in $P$ and $T$, when discretised, leads to a linear eigenvalue problem and so allows the usage of state of the art codes designed for viscous computations with minor modifications. On the other hand the formulation in $p$ leads to a quadratic eigenvalue problem but the scalar nature of the Poincar\'e equation makes it straightforwardly adaptable to oblate spheroidal coordinates as discussed in Appendix~\ref{sec:spheroidal}. 

In the rest of this section, we give the details of the discretisation method assuming the formulation in terms of the Poincar\'e equation. Both formulations are used to obtain the results of Sec.~\ref{sec:results}.

\subsection{Spectral discretisation}
\label{sec:spectraldiscretisation}
\subsubsection{Angular discretisation}
\label{sec:angulardis}
When working in spherical coordinates $\{r,\theta,\phi\}$, the \emph{Partial Differential Equation} (PDE) Eq.~(\ref{eq:Poincare}) can be reduced to a set of coupled \emph{Ordinary Differential Equations} (ODEs) using the method of spherical harmonics decomposition. The pressure field amplitude $p_0$ is written as the following truncated series 
\begin{equation}
p_0(r,\theta,\phi)=\sum_{\ell=0}^L\sum_{m=-\ell}^{\ell}p_{\ell,m}(r)Y_\ell^m(\theta,\phi)~,
\label{eq:Ylmexp}
\end{equation}
where $L$ is an integer chosen as large as possible. The numerical task consists in finding an approximate expression for the radial functions $\{p_{\ell,m}(r)\}$. In the end, the resulting system will be turned into a fully algebraic (matrix) problem by discretisation of the radial direction.

We now review symmetry considerations which allow to decouple the problem further. The second term of Eq.~(\ref{eq:Poincare}) induces a coupling between each harmonic component ($\ell,m$) and its closest neighbours with a degree of the same parity ($\ell\pm2,m$). The origin of the coupling traces back to the presence of the Coriolis force in the momentum equation. Appendix~\ref{sec:appendixSH} gives the analytical expression of this term as well as the last two terms of Eq.~(\ref{eq:bcPoincare}) in spherical coordinates illustrating the case of a spherical container $({\bf\hat{n}}={\bf\hat{r}})$. The boundary condition induces no further coupling of the spherical harmonics components in that case. Moreover, components with different azimuthal $m$ numbers remain uncoupled, an important fact that is carried over to the spheroidal container case (or indeed any axisymmetric container) and allows to solve for modes with different $m$ independently. This classification of modes by their azimuthal $m$ number is no longer applicable for (non-axisymmetric) ellipsoidal containers.

The nature of the coupling of the $\ell$-numbers reflects the decoupling of modes with a pressure profile that is symmetric by reflection across the equatorial plane and those that are anti-symmetric. By using the property, $Y_\ell^m(\theta,\phi)=(-1)^{\ell+m}Y_\ell^m(\pi-\theta,\phi)$, one can show that modes with $(\ell+m)$ {\bf even} are symmetric modes while those with $(\ell+m)$ {\bf odd} are anti-symmetric modes.

Following the above considerations, the layout of the radial functions $p_{\ell,m}(r)$ (which will become the eigenvectors) prior to their discretisation in $r$ reads:
\begin{subequations}
\begin{align}
\left\{p_{|m|,m},p_{|m+2|,m},p_{|m+4|,m},\dots\right\}^\text{T}&&\text{for equatorially sym. modes,}\label{eq:plmsym}\\
\left\{p_{|m+1|,m},p_{|m+3|,m},p_{|m+5|,m},\dots\right\}^\text{T}&&\text{for equatorially antisym. modes.}\label{eq:plmasym}
\end{align}
\end{subequations}

\subsubsection{Radial discretisation}

For the radial discretisation we follow the method proposed by \citet{olver2013} based on polynomial expansion on the (truncated) basis of Chebyshev polynomials for the radial functions and Gegenbauer polynomials for their radial derivatives. Its main advantage compared to the more common collocation methods is that the matrices that result from the discretisation of differential operators are small in size and \emph{sparse}, as opposed to the collocation methods that lead to small but \emph{dense} matrices for equal resolution. Finite difference methods lead to large sparse matrices and do not provide exponential convergence. The sparsity of the resulting matrices in the method of \citet{olver2013} is particularly handy if very high numerical resolution is needed. The only drawback is that it is slightly less straightforward to implement. The use of Chebyshev polynomials is especially convenient due to the possibility to compute coefficients using a \emph{Fast Fourier Transform}. 

We deal with bounding surfaces that deviate slightly from sphericity, with $R$ representing the mean radius of the cavity. In its original form, the spectral method of \citet{olver2013} is designed to deal with differential equations in a single spatial direction $x$, limited to the interval $x\in[-1,1]$. Special care is needed when we consider the radial domain $[0,R]$ of the fluid. One might naively map the radial fluid domain $[0,R]$ to the $[-1,1]$ interval and enforce the appropriate boundary condition (the regularity condition at the centre of coordinates) at $x=-1$. This is however a poor choice because the resulting radial functions are not necessarily compatible with the intrinsic symmetries of the spherical harmonics. A better choice is to extend the fluid domain to $[-R,R]$ and map it to the interval $[-1,1]$ to match the natural domain of the Chebyshev polynomials. Since the spherical harmonics satisfy
\begin{equation}
Y_\ell^m(\theta+\pi,\phi)=(-1)^\ell Y_\ell^m(\theta,\phi),
\end{equation}
the following identity holds for the pressure field $p_0({\bf r})$:
\begin{equation}
p_0(-r,\theta,\phi)=p_0(r,\pi-\theta,\phi+\pi),
\end{equation}
wich implies
\begin{equation}
p_{\ell,m}(-r)=(-1)^\ell p_{\ell,m}(r).
\label{eq:parityplm}
\end{equation}
In our approach each function $p_{\ell,m}(r)$ will be represented as a linear combination of Chebyshev polynomials, $T_k(x)$~: $p_{\ell,m}(r)=\sum_{k=0}^N p_{\ell,m}^kT_k(r/R)$. As the parity of each Chebyshev polynomial can be read from that of its integer index, $k$, the condition Eq.~(\ref{eq:parityplm}) can be enforced by only keeping the coefficients $p_{\ell,m}^k$ that have the correct parity. 

Boundary conditions are enforced by row replacement \citep{olver2013} after expansion on spherical harmonics. 

One important detail that we need to point out is the fact that, when treating a given problem, one should always try to eliminate inverse powers of $r$ from the starting expression. This in order to keep the Chebyshev expansion of any prefactor in the equation as small as possible which greatly reduces the bandwidth of every block matrix \citep{olver2013}. For example, the first step in dealing with Eq.~(\ref{eq:Poincare}) is to multiply it by $r^2$.

The usage of spherical harmonics can be extended to equations written in oblate spheroidal coordinates. An explanation of how to do this is given in Appendix~\ref{sec:spheroidal}.

Fig.\ref{fig:matrices} shows an example of the final set of matrices making up the discretised version of Eqs.~(\ref{eq:Poincare}-\ref{eq:bcPoincare}) for a spherical container using spherical coordinates. The rows dedicated to the enforcement of the boundary condition can be seen at the bottom of each matrix. Fig.~\ref{fig:matricesspheroid} shows the same set of matrices for a spheroidal container using the oblate spheroidal coordinates. Notice that the number of blocks is larger compared to the spherical case of Fig.\ref{fig:matrices}. This illustrates the more extensive coupling between the spherical harmonics components $p_{\ell,m}$. Each individual block also has a wider diagonal representing the increased number of coefficients needed in the polynomial expansion of each component.

\begin{figure}
\begin{tabular}{ccc}
\subfloat[Sphere (spherical coordinates)]{\includegraphics[width=\textwidth]{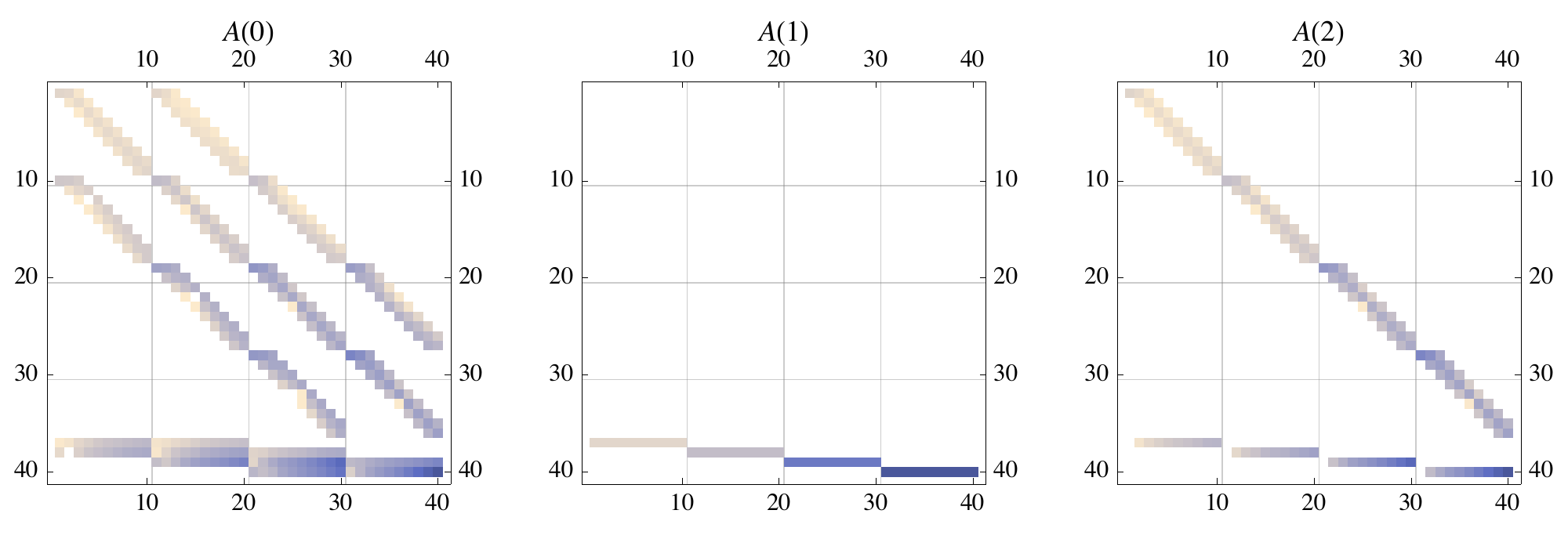}\label{fig:matrices}}\\
\subfloat[Spheroid (oblate spheroidal coordinates)]{\includegraphics[width=\textwidth]{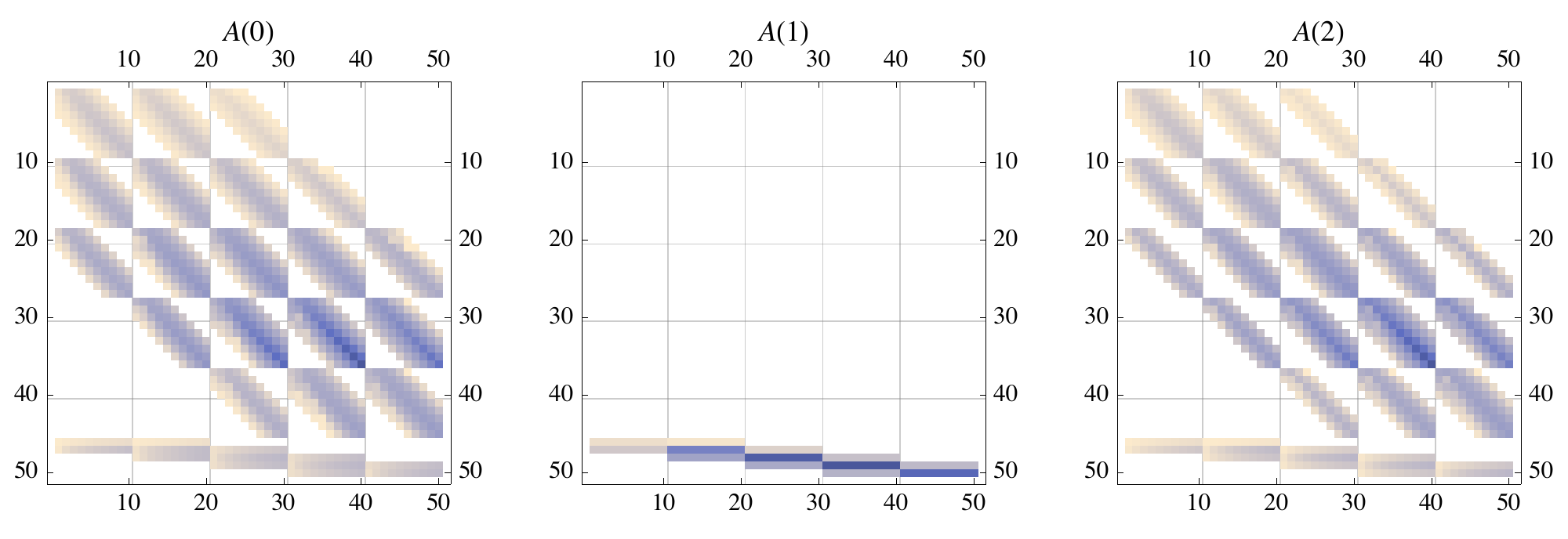}\label{fig:matricesspheroid}}\\
\subfloat[Triaxial ellipsoid (spherical coordinates with series expansion of b.c.)]{\includegraphics[width=\textwidth]{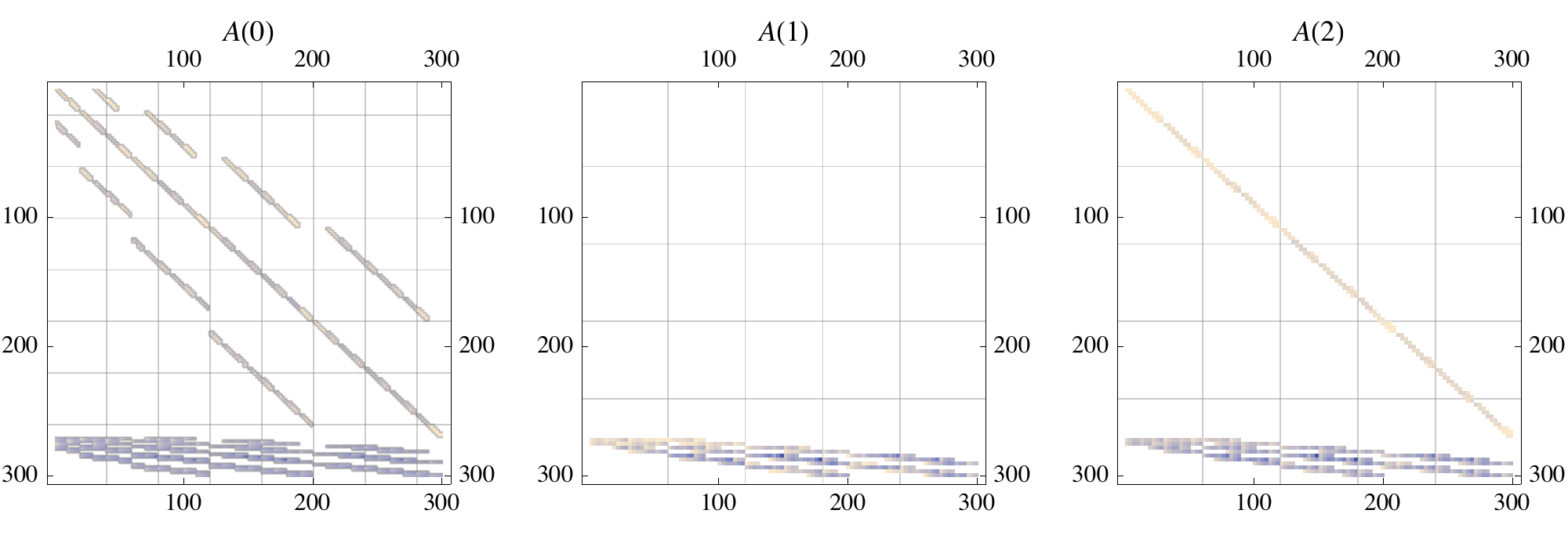}\label{fig:matricestriaxial}}
\end{tabular}
\caption{Matrix representation of the Poincar\'e system Eq.~(\ref{eq:Poincare}-\ref{eq:bcPoincare}) for three types of boundary. The rows dedicated to boundary conditions can be seen at the bottom of each matrix ($N=20, L=10$).\label{fig:matricess}}
\end{figure}

\subsection{Near-spherical boundaries}
\label{sec:nearspherical}
When dealing with non-spherical domains, the use of spherical harmonic decomposition must be adapted. We now present a general way to do so. In essence, it amounts to treat the position of the physical boundary as the result of a series expansion around the sphere. For this reason, this method is only efficient when dealing with \emph{near-spherical boundaries}. The radius of the physical boundary becomes a function of $\theta$ and $\phi$ which we parametrise as
\begin{align}
R(\theta,\phi)&=R_0\left(1+\epsilon(\theta,\phi)\right)\nonumber\\
&=R_0\left(1+\sum_{\ell=0}^\infty\sum_{m=-\ell}^{\ell}\epsilon_{\ell,m}Y_\ell^m(\theta,\phi)\right)~.
\label{eq:bcSH}
\end{align}
The technique will work better if $\epsilon(\theta,\phi)\ll 1$ and the coefficients $\epsilon_{\ell,m}$ are small. The trick is to transform a boundary condition at $r=R(\theta,\phi)$, the \emph{physical boundary}, into a condition on an equivalent spherical domain, the \emph{computational boundary} \citep{smith1974}. For example, suppose that one wishes to impose the Dirichlet boundary condition on a single scalar field $\zeta(r,\theta,\phi)$. At the physical boundary, the condition is simply $\zeta|_{R}=0$. Assuming a small deviation from the spherical boundary, one can write 
\begin{equation}
\zeta|_{R}=\zeta|_{R_0}+\frac{d\zeta}{dr}|_{R_0}(R-R_0)+\mathcal{O}(\epsilon^2)=0~.
\label{eq:dirichlet1storder}
\end{equation}
Using Eq.~(\ref{eq:bcSH}), one finds the following expansion in spherical harmonics
\begin{equation}
\sum_{\ell=0}^L\sum_{m=-\ell}^{\ell}\left(\zeta_{\ell,m}|_{R_0}Y_\ell^m+\sum_{\ell'=0}^L\sum_{m'=-\ell'}^{\ell'}\frac{d\zeta_{\ell,m}}{dr}|_{R_0}\epsilon_{\ell',m'}Y_\ell^mY_{\ell'}^{m'}\right)+\mathcal{O}(\epsilon^2)=0~.
\label{eq:bcphiSH}
\end{equation}
The product of spherical harmonics can then be reduced to a sum using \emph{3-j symbols}.
which will generally consist of $2\ell'+1$ terms featuring spherical harmonics ranging from $Y_{|\ell-\ell'|}^m$ to $Y_{\ell+\ell'}^m$. Those terms effectively couple together these harmonics with $Y_\ell^m$ from the original expression. Imposing the Dirichlet boundary condition -- in its series expansion form -- is therefore equivalent to equating each spherical harmonics component of Eq.~(\ref{eq:bcphiSH}) to zero independently. It is possible to increase the precision of this scheme by keeping higher powers of $\epsilon$ in the Taylor expansion Eq.~(\ref{eq:dirichlet1storder}). This will bring out products of spherical harmonics that include more factors resulting in more extensive coupling in the spherical harmonics.

In general, the expression of the boundary condition may also contain vectors. It is actually the case for Eq.~(\ref{eq:bcPoincare}) which features the pressure gradient, ${\bf\nabla}p$, and the normal vector ${\bf\hat{n}}$. In such cases, the procedure is similar to what we have described, except that we now have to expand both ${\bf \hat{n}}$ and ${\bf\nabla}p$ onto a basis of vector spherical harmonics. The idea remains the same although the handling of symbolic expressions can become quite tedious. In practice, we use \emph{TenGSHui}, a dedicated Mathematica package \citep{trinh2018}, for these manipulations.

The technique described above can be used to deal with ellipsoidal boundaries with small eccentricities. Such boundaries can be parametrised as 
\begin{equation}
x^2+\frac{y^2}{(1-f^2)}+\frac{z^2}{(1-e^2)}=a^2~,
\end{equation}
where $a^2$ stands for the semi-major axis of the ellipsoid in the equatorial plane and $e^2$ and $f^2$ (both taken as $\ll1$) respectively represent the polar and equatorial squared eccentricities in the principal axes of symmetry of the ellipsoid. 

We have already argued that the spherical harmonics components with different azimuthal $m$ numbers decouple in the special case when $f^2=0$ (see Sec.~\ref{sec:angulardis}). In the general (triaxial) case the spherical harmonics expansion will contain terms with different $m$ numbers. The extra amount of coupling greatly increases the computational cost when one increases the angular resolution, \emph{i.e} the upper-bound on the $\ell$ number (see Eq.~(\ref{eq:Ylmexp})). The considerations on the decoupling of the equatorially symmetric and antisymmetric modes, however remains valid with the effect that the shape function, $\epsilon_{\ell,m}$, (see Eq.~(\ref{eq:bcSH})) contains only even $\ell$ and $m$ numbers.\footnote{the explicit expression of each coefficient $\epsilon_{\ell,m}$ is given in Appendix~\ref{sec:seriesell}.} This ensures that modes with even $m$ in an axisymmetric container will have a triaxial counterpart whose expansion contains only even $m$ numbers. The same is true of axisymmetric modes with an odd $m$ number. Fig.~\ref{fig:matricestriaxial} shows the matrices involved in the discretised versions of Eqs.~(\ref{eq:Poincare}-\ref{eq:bcPoincare}) for a triaxial ellipsoid using the method of series expansion of the boundary condition. Notice the larger size of the matrices compared to Fig.~\ref{fig:matrices}-\ref{fig:matricesspheroid} due to the extra-coupling between the coefficients with different $m$ numbers in the boundary conditions.

\subsection{Solver}

We solve the algebraic problem resulting from the discretisation of Eqs.~(\ref{eq:Poincare}-\ref{eq:bcPoincare}) using the SLEPc package \citep{Hernandez2005:SLEPC}. SLEPc is an open-source software built on top of PETSc, another open-source package dedicated to efficiently solve large matrix equations \citep{petsc-efficient,petsc-user-ref,petsc-web-page}. SLEPc is the part that deals with eigenvalue problems. As the matrices involved in our problem can be quite large, it is impractical to solve for the whole eigenspectrum. Instead, we perform a \emph{shift-and-invert} spectral transformation of the original problem which greatly improves the efficiency by limiting the computation of the spectrum to a small region around a given target eigenvalue (the pivot) provided as a guess by the user. Our own usage of SLEPc was greatly inspired by the one described by \citet{vidal2015}. The only difference being that we make use of the built-in PEP solver \citep{Campos2016:SLEPCPEP}. The shift and invert method was also used by \citet{rieutord1997} in a similar context.

\section{Results}
\label{sec:results}

We apply the methods of the previous section to the computation of the inertial modes in an ellipsoid. We start with the spherical case before considering the oblate spheroid and finally the triaxial ellipsoid.

\subsection{Sphere}
\label{sec:resultsphere}
\begin{figure}
\begin{tabular}{ccc}
\subfloat[$\tiny m=1, \bar{\ell}=2$]{\includegraphics[width = 1.6in]{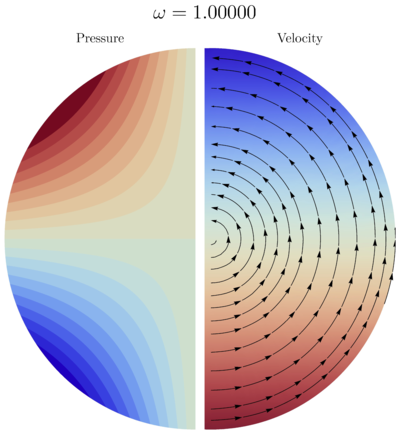}} &
\subfloat[$\tiny m=0, \bar{\ell}=3$]{\includegraphics[width = 1.6in]{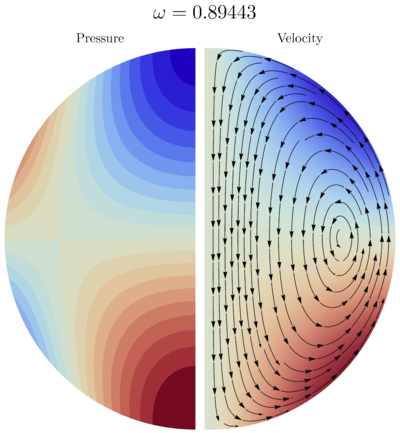}} &
\subfloat[$\tiny m=1, \bar{\ell}=3$]{\includegraphics[width = 1.6in]{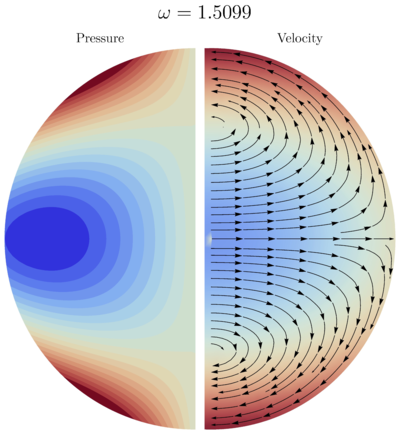}}\\
\subfloat[$\tiny m=1, \bar{\ell}=3$]{\includegraphics[width = 1.6in]{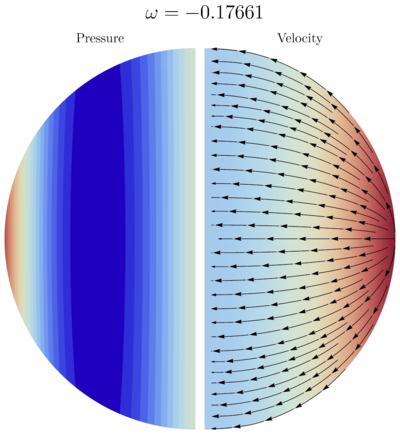}} &
\subfloat[$\tiny m=2, \bar{\ell}=3$]{\includegraphics[width = 1.6in]{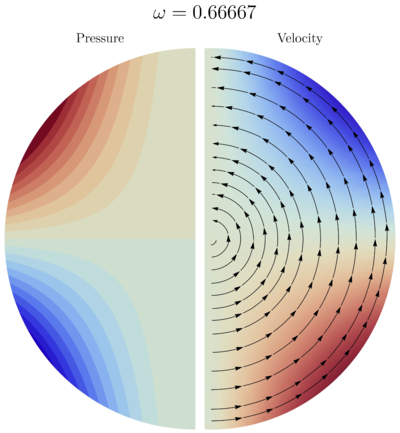}} &
\subfloat[$\tiny m=0, \bar{\ell}=4$]{\includegraphics[width = 1.6in]{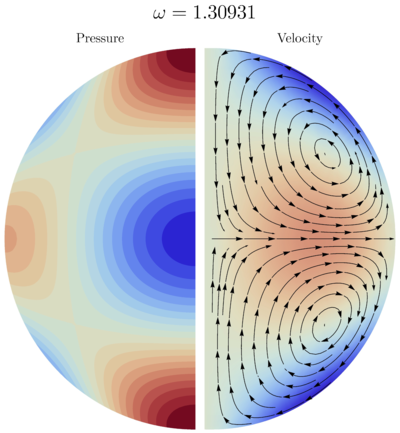}}\\
\subfloat[$\tiny m=1, \bar{\ell}=4$]{\includegraphics[width = 1.6in]{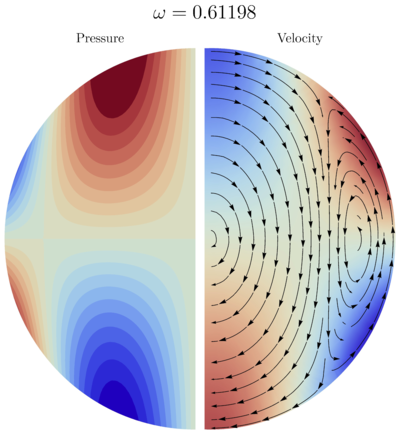}} &
\subfloat[$\tiny m=1, \bar{\ell}=4$]{\includegraphics[width = 1.6in]{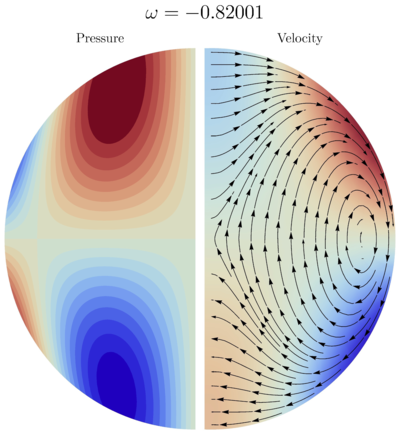}} &
\subfloat[$\tiny m=1, \bar{\ell}=4$]{\includegraphics[width = 1.6in]{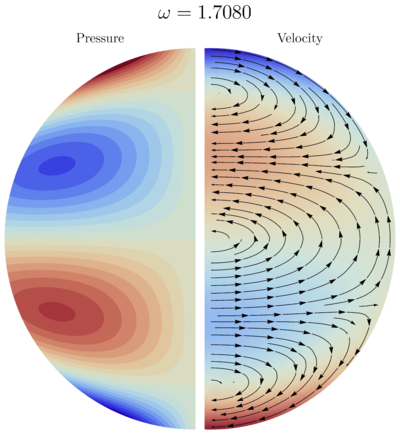}}\\
\subfloat[$\tiny m=2, \bar{\ell}=4$]{\includegraphics[width = 1.6in]{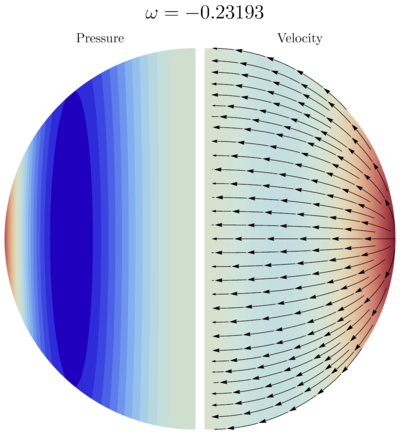}} &
\subfloat[$\tiny m=2, \bar{\ell}=4$]{\includegraphics[width = 1.6in]{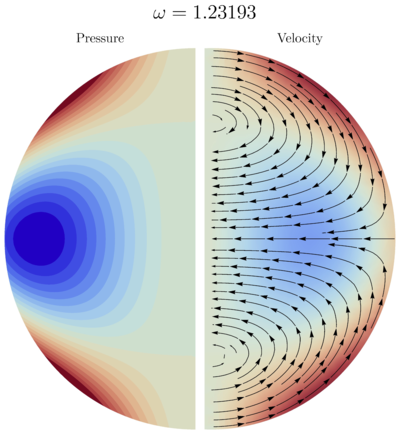}} &
\subfloat[$\tiny m=3, \bar{\ell}=4$]{\includegraphics[width = 1.6in]{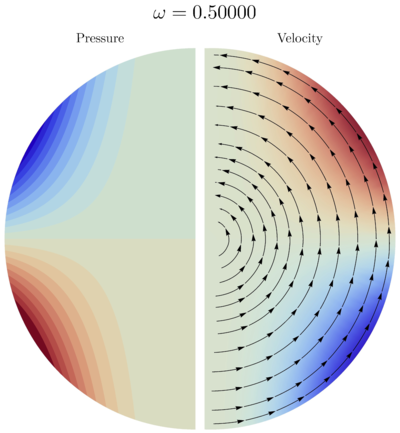}}
\end{tabular}
\caption{Numerical solutions of inviscid inertial modes in a sphere. On each figure, the colours represent the azimuthal component of the velocity on the right and the intensity of the (reduced) pressure inside the cavity on the left. Red corresponds to positive values, blue to negative values. The plots of pressure field are inverted images of the region on the right of each plot. ($N=20$, $L=10$).}
\label{fig:inviscidsphmode}
\end{figure}

Fig.~\ref{fig:inviscidsphmode} shows meridional cuts in the spatial velocity profiles of the first twelve inertial modes in order of increasing spatial complexity. Each mode is identified by its (single) azimuthal $m$ number, its maximum $\ell$ number, $\bar{\ell}$, and its physical frequency, $\omega=2\lambda$. 
The first of these modes which corresponds to $\bar{\ell}=2$ and $m=1$ is of particular importance and is called the \emph{spin-over mode}. In the rotating reference frame, this mode corresponds to a solid body rotation around an axis, itself rotating within the equatorial plane with unit angular frequency.
The computation was carried out using a set of Chebyshev polynomials with maximum degree $N=20$ and a maximum degree of spherical harmonics $L=10$.
The values of the frequencies that we compute agree with the analytical predictions of Eq.~(\ref{eq:Plmomega}) to numerical precision.

\subsection{Oblate spheroid}
\label{sec:resultspheroidal}
\begin{figure}
\begin{tabular}{ccc}
\subfloat[$\tiny m=1, \bar{\ell}=2$]{\includegraphics[width = 1.6in]{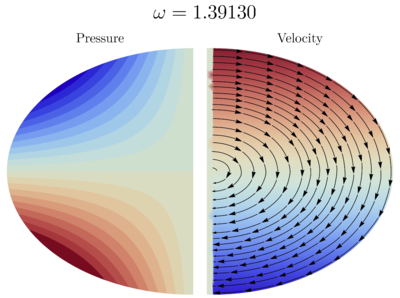}} &
\subfloat[$\tiny m=0, \bar{\ell}=3$]{\includegraphics[width = 1.6in]{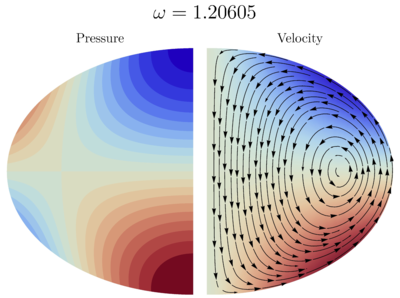}} &
\subfloat[$\tiny m=1, \bar{\ell}=3$]{\includegraphics[width = 1.6in]{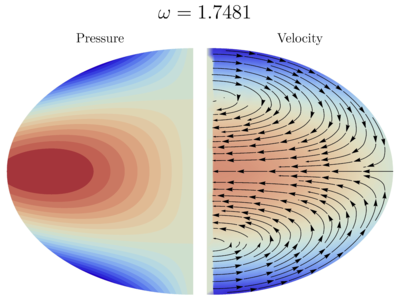}}\\
\subfloat[$\tiny m=1, \bar{\ell}=3$]{\includegraphics[width = 1.6in]{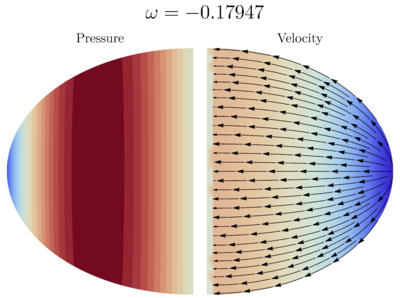}} &
\subfloat[$\tiny m=2, \bar{\ell}=3$]{\includegraphics[width = 1.6in]{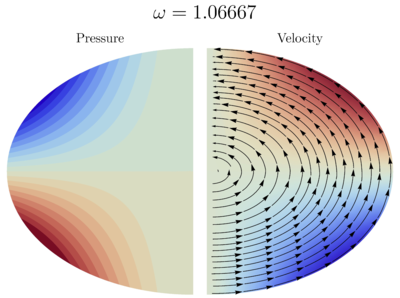}} &
\subfloat[$\tiny m=0, \bar{\ell}=4$]{\includegraphics[width = 1.6in]{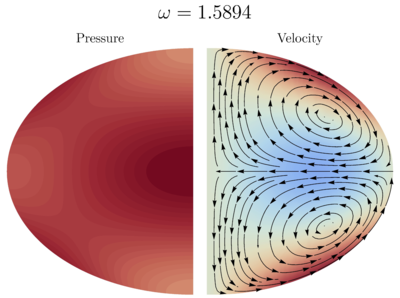}}\\
\subfloat[$\tiny m=1, \bar{\ell}=4$]{\includegraphics[width = 1.6in]{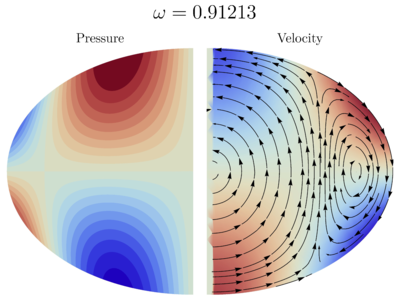}} &
\subfloat[$\tiny m=1, \bar{\ell}=4$]{\includegraphics[width = 1.6in]{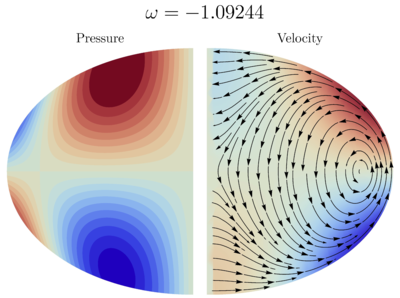}} &
\subfloat[$\tiny m=1, \bar{\ell}=4$]{\includegraphics[width = 1.6in]{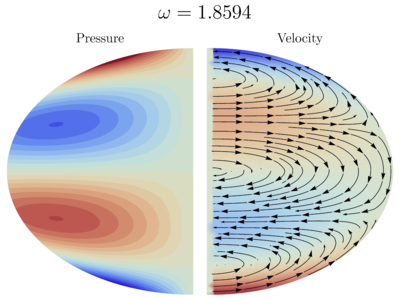}}\\
\subfloat[$\tiny m=2, \bar{\ell}=4$]{\includegraphics[width = 1.6in]{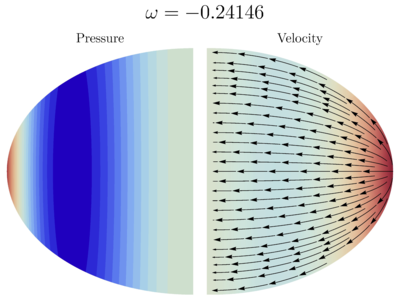}} &
\subfloat[$\tiny m=2, \bar{\ell}=4$]{\includegraphics[width = 1.6in]{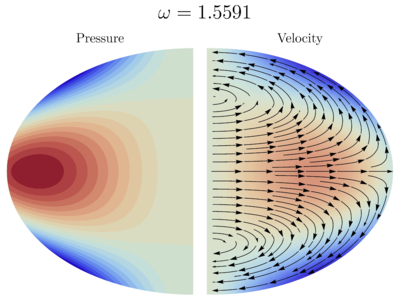}} &
\subfloat[$\tiny m=3, \bar{\ell}=4$]{\includegraphics[width = 1.6in]{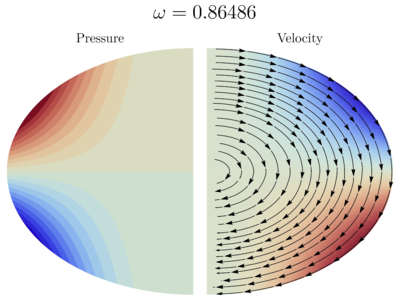}}
\end{tabular}
\caption{Similar to Fig.~\ref{fig:inviscidsphmode} but for an oblate spheroid with squared eccentricity $e^2=\frac{9}{16}$. ($N=40$, $L=20$).}
\label{fig:inviscidmodespheroid}
\end{figure}

Fig.~\ref{fig:inviscidmodespheroid} is analogous to Fig.~\ref{fig:inviscidsphmode} for a spheroidal container of squared eccentricity $e^2=\frac{9}{16}$. The resolution used is $N=20$, $L=10$. The values of the eigenfrequencies again agree with the analytical prediction of Eq.~(\ref{eq:Plmomega}) to numerical precision.
These results were computed using the specially tailored set of oblate spheroidal coordinates (Appendix~\ref{sec:spheroidal}). 

\begin{figure}
\begin{tabular}{ccc}
\subfloat[$\tiny m=1, \bar{\ell}=2$]{\includegraphics[width = 0.5\textwidth]{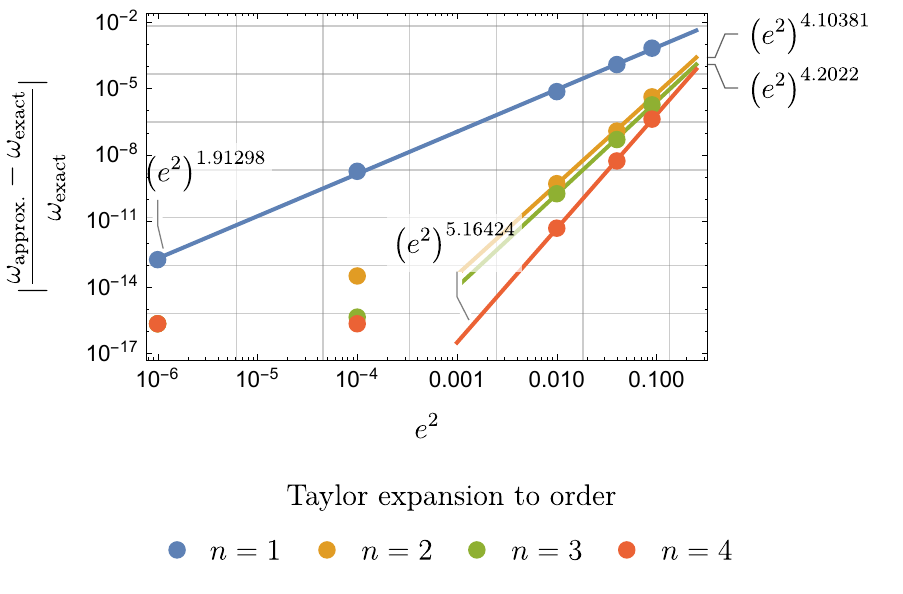}} &
\subfloat[$\tiny m=0, \bar{\ell}=3$]{\includegraphics[width = 0.5\textwidth]{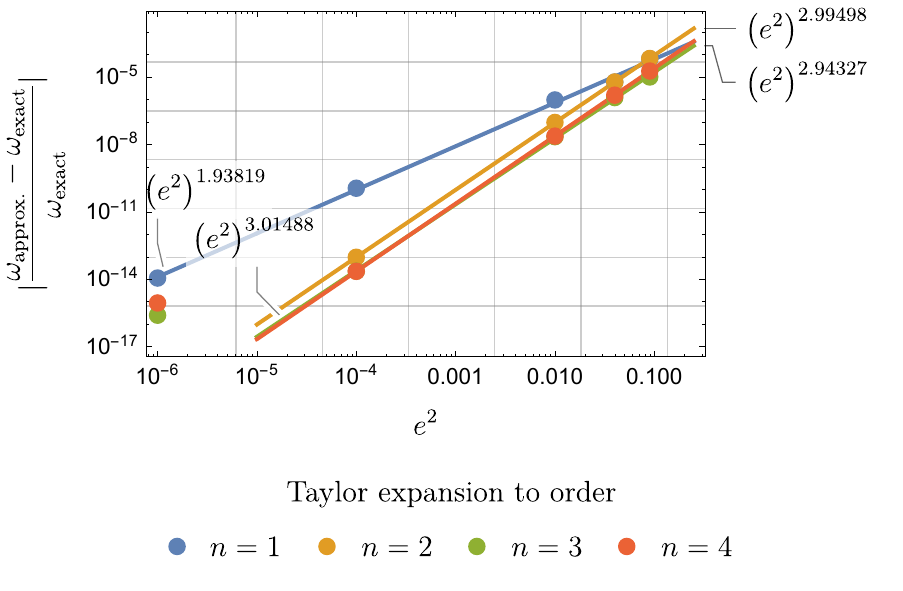}}\\
\subfloat[$\tiny m=1, \bar{\ell}=3$]{\includegraphics[width = 0.5\textwidth]{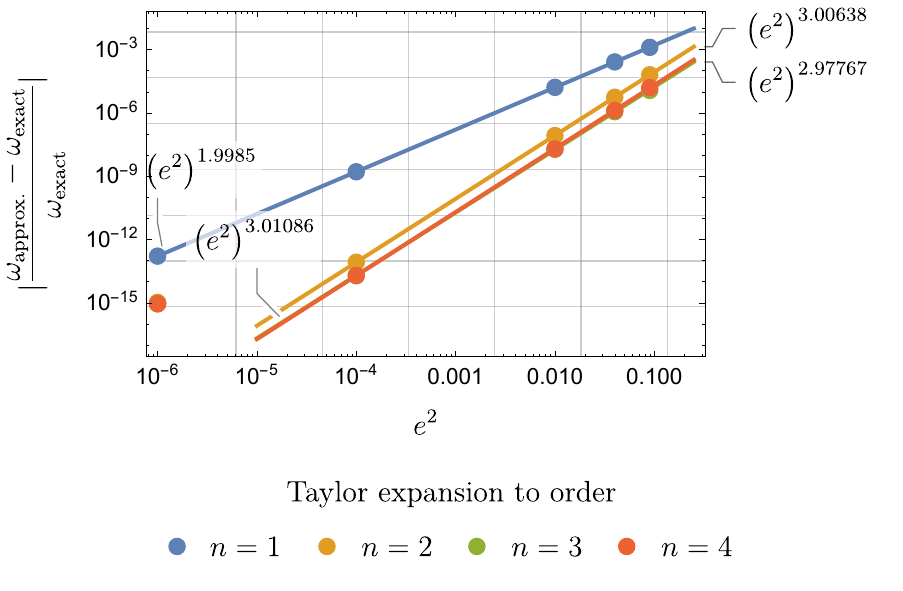}} &
\subfloat[$\tiny m=2, \bar{\ell}=3$]{\includegraphics[width = 0.5\textwidth]{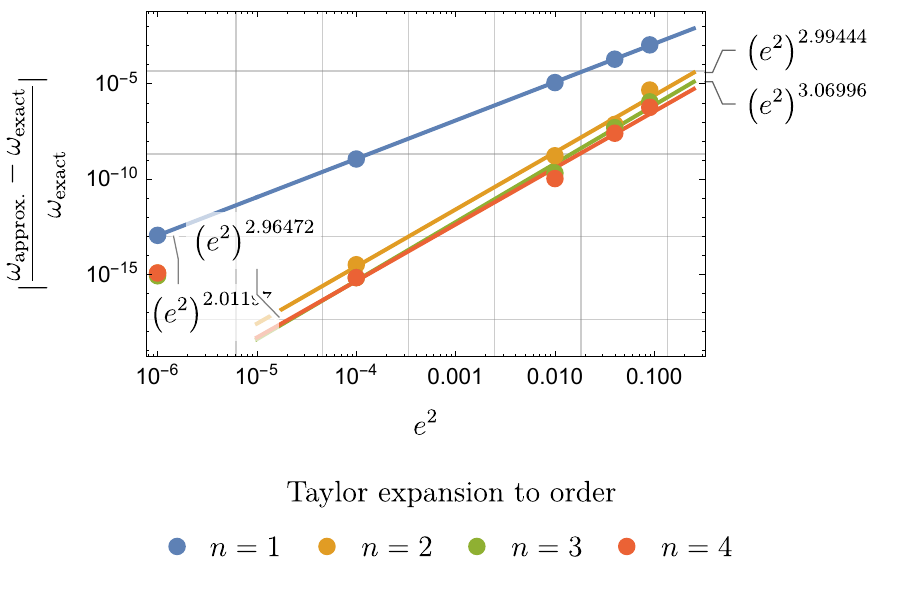}}
\end{tabular}
\caption{{\bf oblate spheroid} -- Comparison of the eigenvalue of the first inertial modes computed using a Taylor expansion of the boundary condition with their analytical values as a function of the squared eccentricity $e^2$ ($N=20$, $L=10$).}
\label{fig:convergence-spheroid}
\end{figure}

The inertial modes of an oblate spheroid can also be computed using the technique of equivalent spherical boundary exposed in Sec.~\ref{sec:nearspherical}. Fig.~\ref{fig:convergence-spheroid} shows the difference between the numerical and analytical eigenvalues for the inertial modes of lowest maximum degree, $\bar{\ell}$. The integer, $n$, represents the maximum order of the Taylor expansion. The error made on the computed eigenfrequencies scales as some near-integer power of the squared eccentricity. We see that the precision on the numerical eigenvalues saturates to second order convergence for all modes except the spin-over ($\bar{\ell}=2, m=1$).
We attribute this to the fact of working in finite arithmetic precision. The simple geometry of the spin-over mode somehow makes this issue less important. Interestingly, the error for $n=2$ for this mode already scales as $\sim(e^2)^4$ so that nothing is gained by using $n=3$.

\subsection{Triaxial ellipsoid}
\label{sec:resulttriaxial}
\begin{figure}
\begin{tabular}{cc}
\subfloat[degree 2, antisymmetric (spin-over)]{\includegraphics[width = 0.5\textwidth]{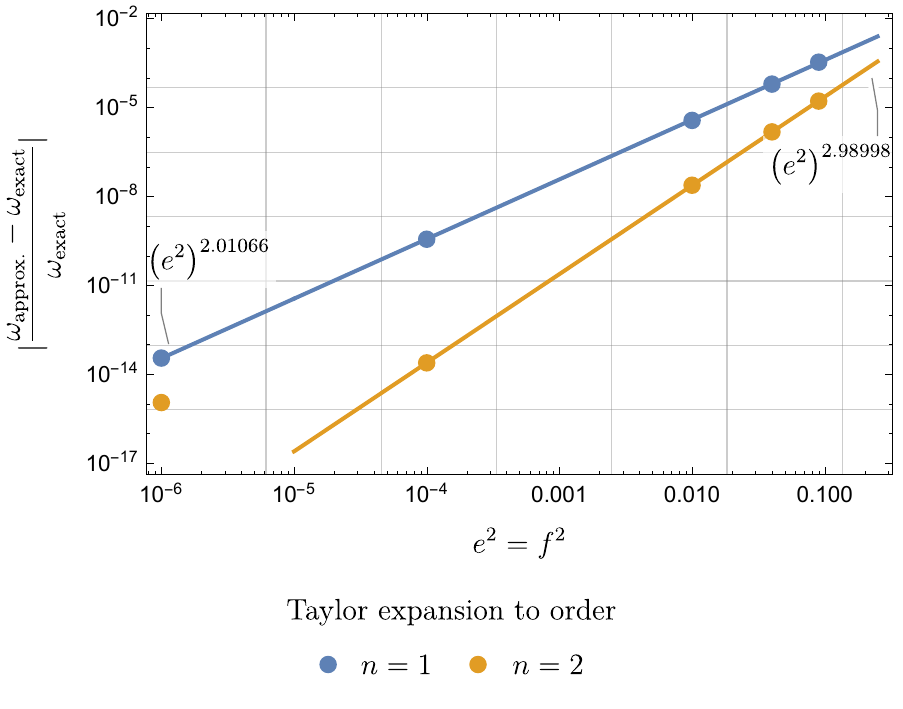}} &
\subfloat[degree 3, symmetric]{\includegraphics[width = 0.5\textwidth]{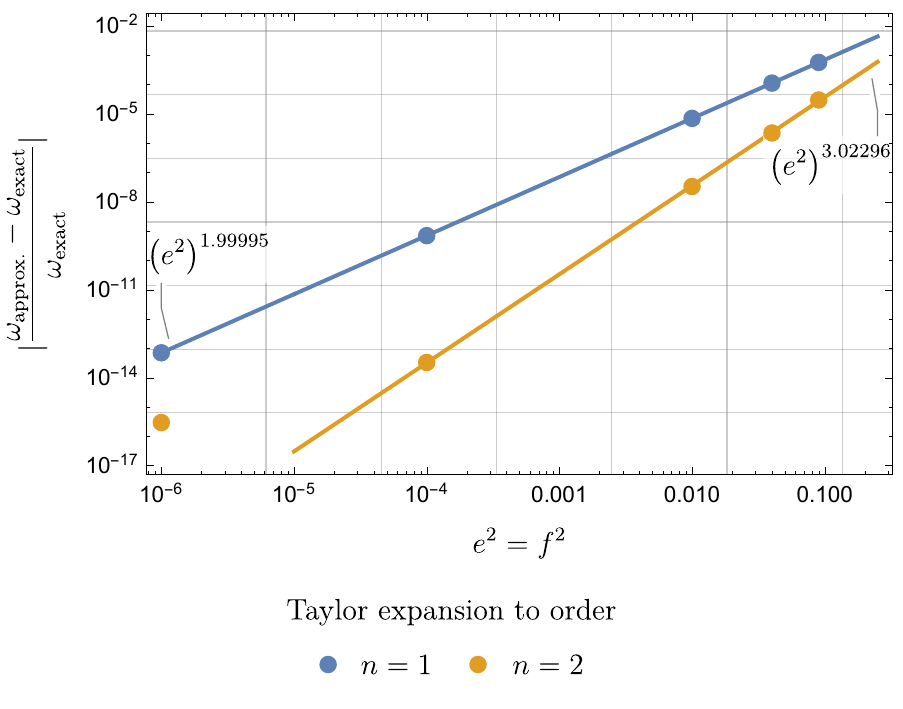}}
\end{tabular}
\subfloat[degree 3, antisymmetric]{\includegraphics[width = 0.5\textwidth]{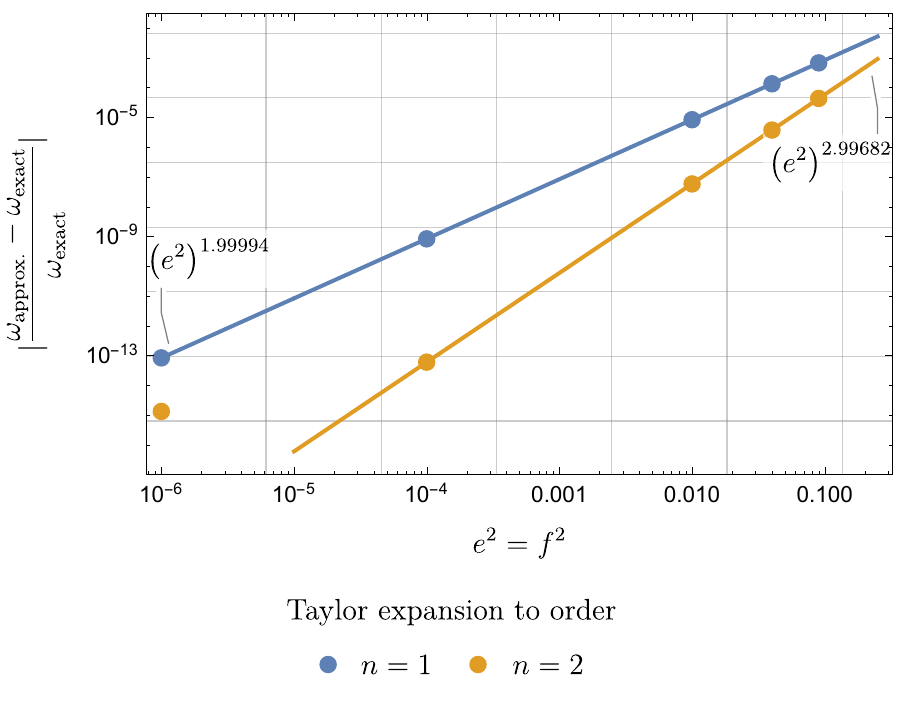}} 
\caption{{\bf triaxial ellipsoid} -- Comparison of the eigenvalue of three inertial modes computed using a Taylor expansion of the boundary condition with their analytical values as a function of the squared eccentricities $e^2 = f^2$ ($N=20$, $L=12$).}
\label{fig:convergence-triaxial-e=f}
\end{figure}

The Taylor expansion of the boundary condition is performed in the two eccentricity parameters $e^2$ and $f^2$ which are considered of the same order of magnitude. Fig.~\ref{fig:convergence-triaxial-e=f} shows the residual error on three eigenvalues in the special case where $e^2=f^2$. The error scales as expected for Taylor expansion to order $n=1$ and $n=2$ respectively. 

\begin{figure}
\begin{tabular}{cc}
\subfloat{\includegraphics[width=0.45\textwidth]{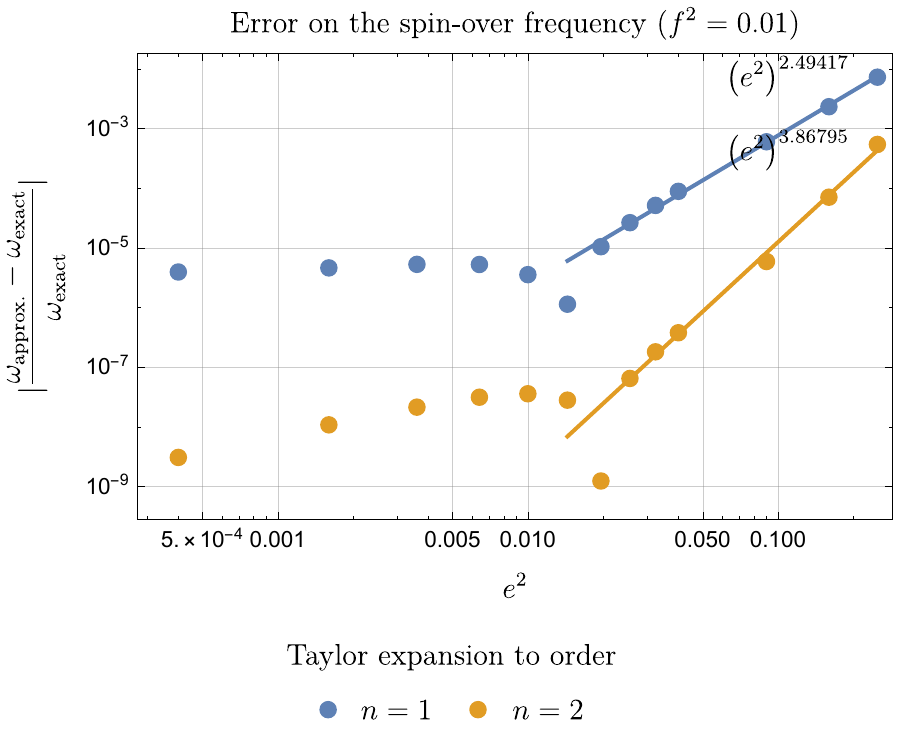}\label{fig:erroromegaellipsoid}}&
\subfloat{\includegraphics[width=0.45\textwidth]{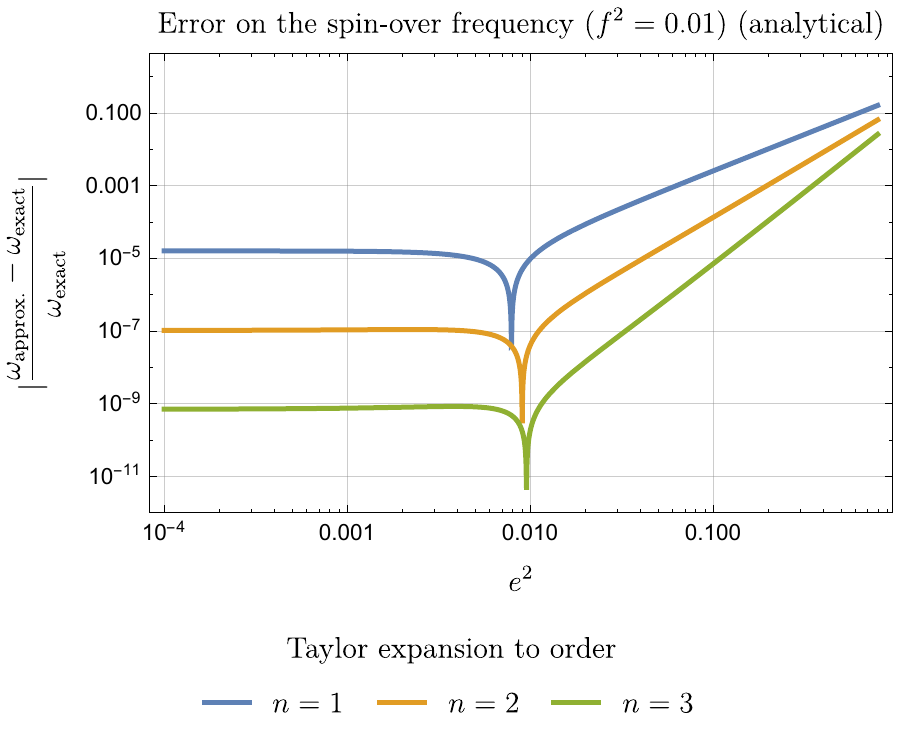}\label{fig:spin-over_triaxial_analytical}}
\end{tabular}
\caption{{\bf triaxial ellipsoid} -- Comparison of the eigenvalue of the spin-over mode. computed using a Taylor expansion of the boundary condition with their analytical values as a function of the squared eccentricities $e^2$ and for $f^2=0.01$. The figure on the right shows the error resulting from the direct Taylor expansion of Eq.~(\ref{eq:spin-over_triaxial}).}
\end{figure}

Fig.~\ref{fig:erroromegaellipsoid} shows the error on the evaluation of the spin-over frequency as a function of $e^2$, this time setting $f^2=0.01$. The dip around $e^2\sim0.01$ corresponds to the point where the error changes sign. This feature is also present on Fig.~\ref{fig:spin-over_triaxial_analytical} which shows the error made by directly expanding the analytical formula of Eq.~(\ref{eq:spin-over_triaxial}) in powers of $e^2$ and $f^2$. Decreasing the value of $e^2$ below the threshold $e^2=f^2$ has no effect on the accuracy of the solution which becomes limited by the value of $f^2$, hence the saturation observed on the left side of both graphs. Above the threshold, one recovers the expected power laws in $e^2$ indicating the good numerical convergence properties of the method.
%
%
%
%

\section{Discussion}
\label{sec:discussion}

We have presented a new numerical method to compute the inviscid inertial modes of a rotating near-spherical container. This is based on the fully spectral discretisation of the angular and radial directions using spherical harmonics in the angular directions and Chebyshev as well as Gegenbauer polynomials in the radial direction. We have shown how the method can be used to solve the Poincar\'e equation and the no penetration condition simultaneously as an algebraic polynomial eigenvalue problem and we have employed it to recover the inertial modes of a sphere and an oblate spheroid numerically with an accuracy limited by machine precision only (Sec.~\ref{sec:resultsphere} and \ref{sec:resultspheroidal}). We have also shown how the method of equivalent spherical boundary introduced by \citet{smith1974} can be used to compute the inertial modes inside a boundary that deviates only slightly from a sphere. We substantiated this technique with an analysis of its numerical convergence for the inertial modes of lowest degree both in an oblate axisymmetric ellipsoid and a triaxial ellipsoid with small eccentricities (Sec.~\ref{sec:resultspheroidal} and \ref{sec:resulttriaxial}). 

The demonstration of the well posedness and numerical convergence of the method exposed in this paper in the context of inviscid inertial modes is an important first step towards its application to other problems for which there exists no analytical solution. It is not limited to the study of the sole Poincar\'e equation and can, in principle, be applied to the resolution of any system of differential equations in near-spherical geometry. Such problems are ubiquitous to the fields of planetology and geophysical/astrophysical fluid dynamics. Future foreseen applications include the study of the impact of inertial modes on the global rotation of a planet via coupling with the Liouville equation of rotational dynamics. This will be the subject of a future paper by the authors \citep{triana2018}. This study also investigates the effect of viscosity. The effects of stratification and magnetisation of the liquid core are also among possible applications. Such studies are natural extensions to the formalism of the present paper via inclusion of the equations for heat diffusion and magnetic induction. With minor modifications, the spectral discretisation described here can also accommodate multi-layered physical configurations. This would allow to couple the dynamics of the liquid core to that of a visco-elastic and self gravitating mantle with topography. These studies are long-terms endeavours that are currently being investigated by the authors.


%
\section*{Acknowledgement}
The authors would like to thank J.~Vidal and N.~Schaeffer for the useful discussions in the first stage of development of the method which lead to the present paper. We would also like to thank S.~Olver and the PETSc and SLEPc support teams for their help with technical details in the making of this work.
The research leading to these results has received funding from the European Research Council (ERC) under the European Union's Horizon 2020 research and innovation programme (Advanced Grant agreement No~670874).

\bibliographystyle{abbrvnat}
\bibliography{biblio}

\begin{appendix}
\newpage

\section{Analytical solutions}
\label{sec:analytical}
The idea is to introduce a new (frequency-dependent) set of coordinates $\{u,v,w\}$
\begin{subequations}
\begin{align}
x&=X(u,v,w)\\
y&=Y(u,v,w)\\
z&=\beta Z(u,v,w)~.
\end{align}
\end{subequations}
with the rescaling factor
\begin{equation}
\beta=\sqrt{\frac{1-\lambda^2}{\lambda^2}}\qquad\text{where}\qquad\lambda=\frac{\omega}{2\Omega}
\label{eq:beta}
\end{equation}
so as to turn the Poincar\'e equation (Eq.~(\ref{eq:Poincare})) into a Laplace equation:
\begin{align}
\nabla^2 p_0-\frac{1}{\lambda^2}({\bf \hat{z}}\cdot{\bf \nabla})^2 p_0&=\left(\frac{\partial^2 p_0}{\partial x^2}\right)_{y,z}+\left(\frac{\partial^2 p_0}{\partial y^2}\right)_{x,z}-\beta^2\left(\frac{\partial^2 p_0}{\partial z^2}\right)_{x,y}\\
&=\left(\frac{\partial^2 p_0}{\partial X^2}\right)_{Y,Z}+\left(\frac{\partial^2 p_0}{\partial Y^2}\right)_{X,Z}+\left(\frac{\partial^2 p_0}{\partial Z^2}\right)_{X,Y}\\
&=0
\label{eq:Poincarepartial}
\end{align}
The set of coordinates $\{u,v,w\}$ should be chosen so that the Laplace equation is separable and the no-penetration boundary condition (Eq.~(\ref{eq:bcPoincare})) is easy to implement.

From the implicit expression for the general ellipsoid in $\mathbb{R}^3$
\begin{equation}
\Phi\equiv\frac{x^2}{a^2}+\frac{y^2}{b^2}+\frac{z^2}{c^2}-1=0~,
\end{equation}
the (unnormalised) normal vector ${\bf n}\sim{\bf \nabla}\Phi$ at a surface point $x{\bf \hat{x}}+y{\bf \hat{y}}+z{\bf \hat{z}}$ writes
\begin{equation}
{\bf n}=\frac{x}{a^2}{\bf \hat{x}}+\frac{y}{b^2}{\bf \hat{y}}+\frac{z}{c^2}{\bf \hat{z}}~.
\end{equation}
Therefore, the boundary condition Eq.~(\ref{eq:bcPoincare}) in Cartesian coordinates reads
\begin{equation}
-\left(\frac{i y\lambda}{b^2}+\frac{x\lambda^2}{a^2}\right)\left(\frac{\partial p_0}{\partial x}\right)_{y,z}+\left(\frac{i x\lambda}{a^2}-\frac{y\lambda^2}{b^2}\right)\left(\frac{\partial p_0}{\partial y}\right)_{x,z}+\left(\frac{z}{c^2}-\frac{z\lambda^2}{c^2}\right)\left(\frac{\partial p_0}{\partial z}\right)_{x,y}=0~.
\label{eq:3axebccart}
\end{equation}
evaluated at a surface point $x{\bf \hat{x}}+y{\bf \hat{y}}+z{\bf \hat{z}}$. To rewrite the boundary condition Eq.~(\ref{eq:3axebccart}) in terms of the new set of coordinates $\{u,v,w\}$, we compute the Jacobian of the transformation of coordinates
\begin{equation}
[{\bf J}]=\begin{bmatrix}
    \left(\dfrac{\partial X}{\partial u}\right)_{v,w} & \left(\dfrac{\partial X}{\partial v}\right)_{u,w} & \left(\dfrac{\partial X}{\partial w}\right)_{u,v} \\
    \left(\dfrac{\partial Y}{\partial u}\right)_{v,w} & \left(\dfrac{\partial Y}{\partial v}\right)_{u,w} & \left(\dfrac{\partial Y}{\partial w}\right)_{u,v} \\
    \beta\left(\dfrac{\partial Z}{\partial u}\right)_{v,w} & \beta\left(\dfrac{\partial Z}{\partial v}\right)_{u,w} & \beta\left(\dfrac{\partial Z}{\partial w}\right)_{u,v}
\end{bmatrix}
\end{equation}
and deduce
\begin{equation}
\begin{bmatrix}
    \left(\dfrac{\partial p_0}{\partial x}\right)_{y,z} &
    \left(\dfrac{\partial p_0}{\partial y}\right)_{x,z} &
    \left(\dfrac{\partial p_0}{\partial z}\right)_{x,y}
\end{bmatrix}=
\begin{bmatrix}
    \left(\dfrac{\partial p_0}{\partial u}\right)_{v,w} &
    \left(\dfrac{\partial p_0}{\partial v}\right)_{u,w} &
    \left(\dfrac{\partial p_0}{\partial w}\right)_{u,v}
\end{bmatrix}\cdot[{\bf J}]^{-1}
\end{equation}

We now show that such a convenient set of coordinates can be found in spherical, spheroidal, and ellipsoidal geometries. Note that in the sequel we implicitly assume $0<\lvert\lambda\rvert<1$. Sec.~2.7 of \citet{greenspan1968} shows that all inertial modes in general bounded geometries have a half-frequency $-1\leq\lambda\leq 1$, that $\lambda=\pm 1$ is not an eigenfrequency, and that $\lambda=0$ is an eigenfrequency with infinite multiplicity.

\subsection{Sphere and oblate spheroid}
\begin{figure}
\includegraphics[width=0.66\textwidth]{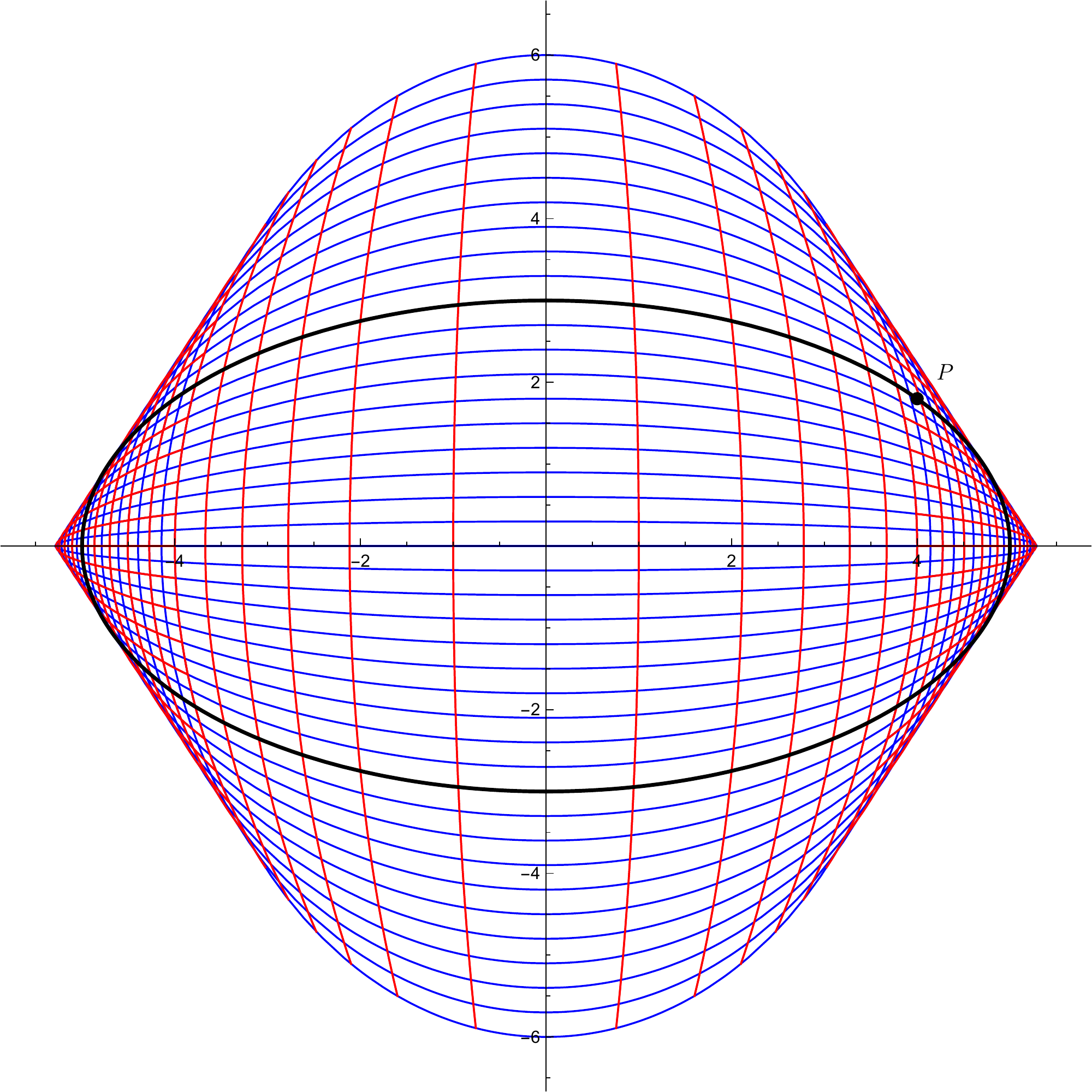}
\caption{2D projection of the set of bispheroidal coordinates (Eqs.~(\ref{eq:bispheroidal})) for $\lambda=\frac{1}{2}$ in the $xz$-plane ($\phi=0$). Blue curves are level surfaces of constant $\xi$, red curves are level surfaces of constant $\mu$. The point $P$ belongs to the physical ellipsoid which is shown as the thick black curve.}
\label{fig:bispheroidalcoordinates}
\end{figure}

The appropriate set of coordinates in this case is the (frequency-dependent) set of \emph{bi-spheroidal} coordinates $\{\xi,\mu,\phi\}$\footnote{Note that these are different from the orthogonal \emph{oblate spheroidal coordinates} presented in Appendix~\ref{sec:spheroidal}.}, which are defined, in terms of the Cartesian coordinates, as
\begin{subequations}
\begin{align}
x&=k\sqrt{1-\xi^2}\sqrt{1-\mu^2}\cos\phi\\
y&=k\sqrt{1-\xi^2}\sqrt{1-\mu^2}\sin\phi\\
z&=\beta k\xi\mu~.
\end{align}
\label{eq:bispheroidal}
\end{subequations}
with $\beta$ as previously (Eq.~(\ref{eq:beta})) and
\begin{equation}
k=\sqrt{a^2+\frac{c^2}{\beta^2}}=a\sqrt{\frac{1-e^2\lambda^2}{1-\lambda^2}},
\label{eq:k}
\end{equation}
where $0<\xi<1$, $-\xi<\mu<\xi$, and $0<\phi<2\pi$. Note that these coordinates only map a bounded domain shaped as a pair of \emph{cuberdons} joined along their base (corresponding to $\xi=\mu$). For convenience, we allow a double covering of this domain by extending $\mu$ to $-1<\mu<1$. Then, the level surfaces of $\xi$ are (prolate or oblate) spheroids, those of $\mu$ are half-spheroids, and those of $\phi$ are vertical half-planes. Fig.~\ref{fig:bispheroidalcoordinates} gives a graphical representation of these coordinates in the $xz$-plane. Notice how this system of coordinates is \emph{non-orthogonal}.

Under the above transformation, the Poincar\'e equation takes the form of a Laplace equation, of which the general solutions are solid bi-spheroidal harmonics,
\begin{equation}
p=\sum_{\ell=0}^{+\infty}\sum_{m=-\ell}^{+\ell}p_{\ell m}\mathrm{P}_\ell^m(\xi)\mathrm{P}_\ell^m(\mu)e^{i m\phi}~,
\label{eq:pPlm}
\end{equation}
where the $p_{\ell m}$ are constants and $\mathrm{P}_\ell^m$ is the degree-$\ell$ order-$m$ associated Legendre function of the first kind. Regularity conditions prohibit the appearance of associated Legendre functions of the second kind $\mathrm{Q}_\ell^m$.

One interest of bi-spheroidal coordinates is that the surface of the oblate spheroid is a level surface of $\xi$,
\begin{equation}
\frac{x^2+y^2}{a^2}+\frac{z^2}{c^2}=1\qquad\Leftarrow\qquad\xi=\lvert\lambda\rvert\sqrt{\frac{1-e^2}{1-e^2\lambda^2}},
\end{equation}
which makes it easier to impose the no-penetration boundary condition. The latter requires the normal velocity to vanish, and can be expressed in terms of the pressure as Eq.~(\ref{eq:3axebccart}), written in bi-spheroidal coordinates. Plug Eq.~(\ref{eq:pPlm}) into Eq.~(\ref{eq:3axebccart}):
\begin{multline}
\sum_{\ell=0}^{+\infty}\sum_{m=-\ell}^{+\ell}p_{\ell m}\Biggl((1-\lambda^2)\operatorname{sign}(\lambda){\mathrm{P}_{\ell}^{m}}'\left(\lvert\lambda\rvert\sqrt{\frac{1-e^2}{1-e^2\lambda^2}}\right)\\
-m\sqrt{(1-e^2)(1-e^2\lambda^2)}\mathrm{P}_{\ell}^{m}\left(\lvert\lambda\rvert\sqrt{\frac{1-e^2}{1-e^2\lambda^2}}\right)\Biggr)\mathrm{P}_\ell^m(\mu)e^{i m\phi}=0
\end{multline}
The functions $\mathrm{P}_\ell^m(\mu)e^{i m\phi}$ form a basis of surface spheroidal harmonics over the surface of the oblate spheroid. The boundary condition therefore leaves the coefficients $p_{\ell m}$ uncoupled, another advantage of bi-spheroidal coordinates. By orthogonality of the surface spheroidal harmonics, each harmonic coefficient has to vanish. The frequencies of the inertial modes in an oblate spheroid correspond to those values of $\lambda$ or $\omega$ that allow non-zero values for $p_{\ell m}$, i.e. one of the multiplicative factors vanishes. After some rearrangement,
\begin{equation}
{\mathrm{P}_{\ell}^{m}}'\left(\lambda\sqrt{\frac{1-e^2}{1-e^2\lambda^2}}\right)-\frac{m\sqrt{(1-e^2)(1-e^2\lambda^2)}}{1-\lambda^2}\mathrm{P}_{\ell}^{m}\left(\lambda\sqrt{\frac{1-e^2}{1-e^2\lambda^2}}\right)=0~.
\label{eq:Plmomega}
\end{equation}

The sphere corresponds to the special case where $e^2=0$, in which case one recovers Eq.~(2.12.8) of \citet{greenspan1968}.

The so-called \emph{planetary modes} \citep{rieutord2014}, or \emph{r-modes}, are a special family of solutions that satisfy $\ell=\lvert m\rvert+1$. Inserting the latter into Eq.~(\ref{eq:Plmomega}) reduces it to an equation of the first degree in $\lambda$ with solution \citep{zhang2004},
\begin{equation}
\lambda=\frac{\operatorname{sign}(m)}{1+\lvert m\rvert(1-e^2)}~.
\label{eq:planetaryomega}
\end{equation}
The \emph{spin-over mode} is the simplest member of this family, corresponding to $m=\pm 1$ and $\ell=2$.

\subsection{Triaxial ellipsoid}
\begin{figure}
\begin{tabular}{cc}
\subfloat[$xz$-plane]{\includegraphics[width=0.6\textwidth]{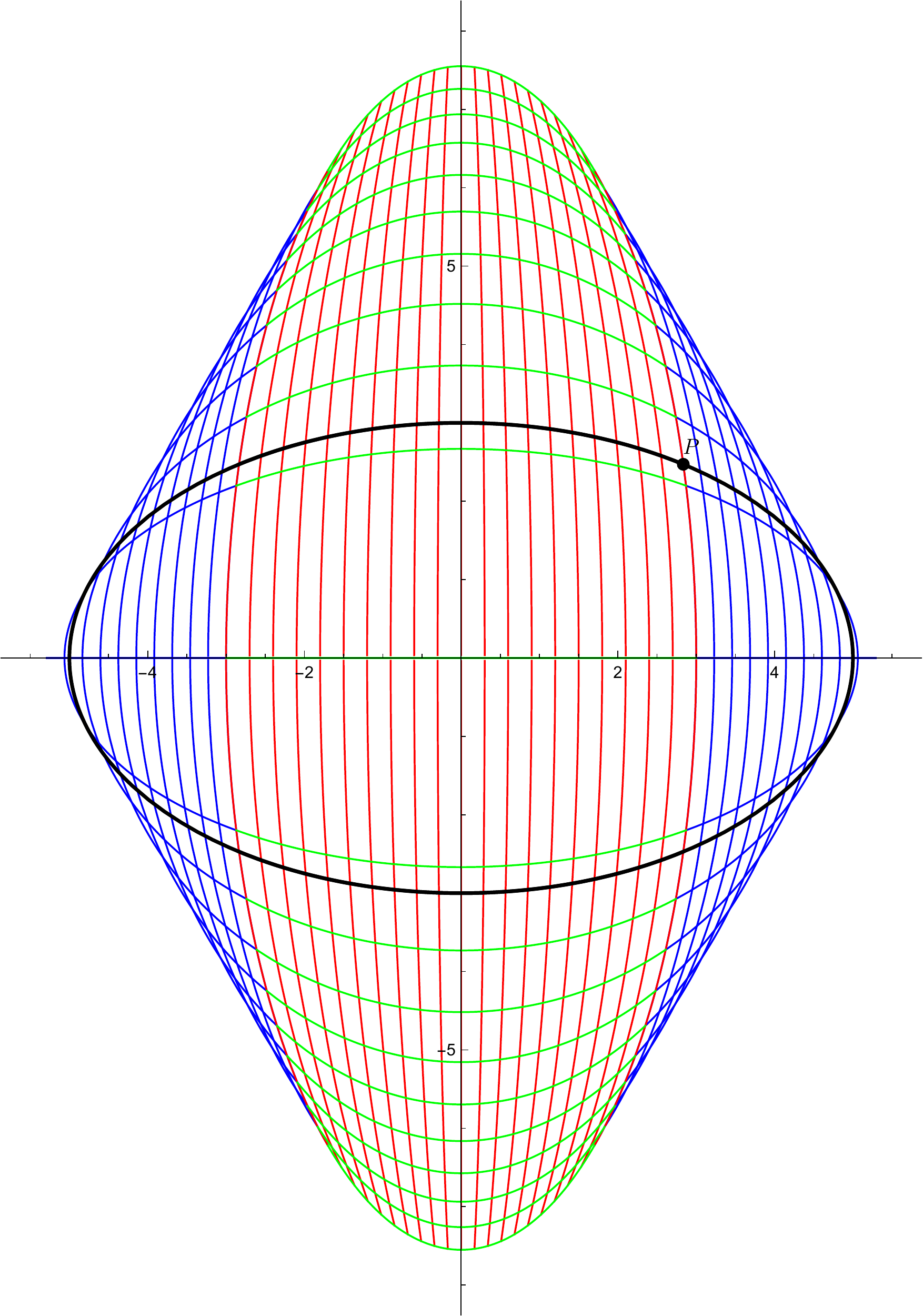}}\\
\subfloat[$xy$-plane]{\includegraphics[width=0.6\textwidth]{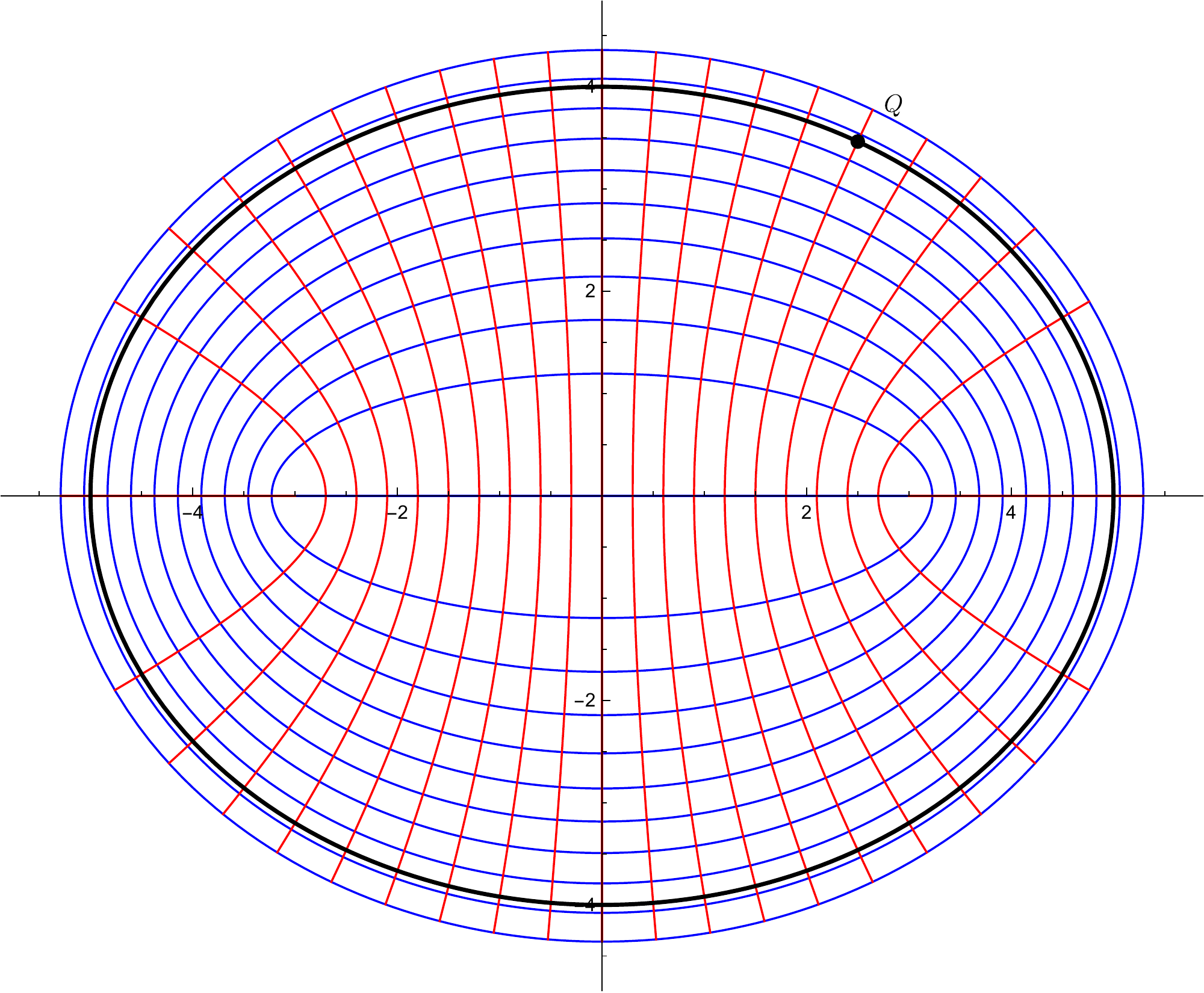}}
\end{tabular}
\caption{2D projections of the set of biellipsoidal coordinates (Eqs.~(\ref{eq:biellipsoidal})) for $\lambda=\frac{1}{2}$ in the $xz$-plane ($\nu$ or $\mu=h$) and $yz$-plane ($\nu$ or $\mu=k$) respectively. Blue curves are level surfaces of constant $\mu$, red curves are level surfaces of constant $\nu$, green curves are level surfaces of constant $\rho$. The point $P$ and $Q$ belong to the physical ellipsoid which is shown as the thick black curve.}
\label{fig:biellipsoidalcoordinates}
\end{figure}

The appropriate set of coordinates in this case is the (frequency-dependent) set of \emph{bi-ellipsoidal} coordinates $\{\rho,\mu,\nu\}$, which are defined, in terms of the Cartesian coordinates, as
\begin{subequations}
\begin{align}
x&=\pm\frac{\rho\mu\nu}{hk}\\
y&=\pm\frac{\sqrt{\rho^2-h^2}\sqrt{\mu^2-h^2}\sqrt{h^2-\nu^2}}{h\sqrt{k^2-h^2}}\\
z&=\pm\frac{\beta\sqrt{k^2-\rho^2}\sqrt{k^2-\mu^2}\sqrt{k^2-\nu^2}}{k\sqrt{k^2-h^2}}~.
\end{align}
\label{eq:biellipsoidal}
\end{subequations}
with $\beta$ and $k$ as previously (Eqs.~(\ref{eq:beta}-\ref{eq:k})) and
\begin{equation}
h=\sqrt{a^2-b^2}=a\sqrt{1-f^2},
\end{equation}
where $h<\rho<k$, $h<\mu<\rho$, and $0<\nu<h$. Note that these coordinates only map one octant of a bounded domain shaped as a pair of cuberdons joined along their base (corresponding to $\rho=\mu$), hence the $\pm$ signs. For convenience, we allow a double covering of this domain by extending $\mu$ to $h<\mu<k$. Then, the level surfaces of $\rho$ and $\mu$ are (sections of) ellipsoids, those of $\nu$ are (sections of) 1-sheeted hyperboloids. Fig.~\ref{fig:biellipsoidalcoordinates} gives a graphical representation of these coordinates in the $xz$ and $xy$ planes. Notice how this system of coordinates is \emph{non-orthogonal}.

Under the above transformation, the Poincar\'e equation takes the form of a Laplace equation, of which the general solutions are solid bi-ellipsoidal harmonics,
\begin{equation}
p=\sum_{n=0}^\infty\sum_{p=1}^{2n+1}c_{np}\mathrm{E}_{n,p}(\rho)\mathrm{E}_{n,p}(\mu)\mathrm{E}_{n,p}(\nu)~,
\label{eq:pEnp}
\end{equation}
where the $c_{np}$ are constants and $\mathrm{E}_{n,p}$ is the degree-$n$ index-$p$ Lam\'e function of the first kind\footnote{see \emph{e.g.} pp.1304-1309 of \citet{morse1953} for reference.} (slightly modified to ensure that it always evaluates to a real number). Regularity conditions prohibit the appearance of Lam\'e functions of the second kind $\mathrm{F}_{n,p}$.

One interest of bi-ellipsoidal coordinates is that the surface of the triaxial ellipsoid is a level surface of $\rho$,
\begin{equation}
\frac{x^2}{a^2}+\frac{y^2}{b^2}+\frac{z^2}{c^2}=1\qquad\Leftarrow\qquad\rho=a,
\end{equation}
which makes it easier to impose the no-penetration boundary condition. The latter requires the normal velocity to vanish, and can be reexpressed in terms of the pressure as Eq.~(\ref{eq:3axebccart}), written in bi-ellipsoidal coordinates. Plug Eq.~(\ref{eq:pEnp}) into Eq.~(\ref{eq:3axebccart}):
\tiny
\begin{align}
\sum_{n=0}^\infty\sum_{p=1}^{2n+1}c_{np}\Biggl(
a(\mu^2-\nu^2)\lambda\frac{\sqrt{(1-f^2)(1-e^2\lambda^2)(1-e^2\lambda^2-f^2(1-\lambda^2))}}{\sqrt{(a^2f^2-\mu^2)(a^2f^2-\nu^2)}}\mathrm{E}_{n,p}'(a)\mathrm{E}_{n,p}(\mu)\mathrm{E}_{n,p}(\nu)+\nonumber\\
\mathrm{E}_{n,p}(a)\left(\nu  \left(1-e^2 \lambda ^2-\frac{\mu ^2 \left(1-\lambda ^2\right)}{a^2}\right)\mathrm{E}_{n,p}'(\mu)\mathrm{E}_{n,p}(\nu)-\mu  \left(1-e^2 \lambda ^2-\frac{\nu ^2 \left(1-\lambda ^2\right)}{a^2}\right)\mathrm{E}_{n,p}(\mu)\mathrm{E}_{n,p}'(\nu)\right)\Biggr)
=0~.
\label{eq:Enplambda}
\end{align}
\normalsize
The functions $\mathrm{E}_{n,p}(\mu)\mathrm{E}_{n,p}(\nu)$ form a basis of surface ellipsoidal harmonics over the surface of the triaxial ellipsoid. The boundary condition therefore only slightly couples the coefficients $c_{np}$, another advantage of bi-ellipsoidal coordinates. The second term of Eq.~(\ref{eq:Enplambda}) only mixes terms of the same degree and symmetry. Indeed, the second-line terms can be rewritten as a sum of surface bi-ellipsoidal harmonics \citep{hough1895}. There is, however, no general expression and each degree and symmetry has to be considered individually. By orthogonality of the surface ellipsoidal harmonics, each harmonic coefficient has to vanish. This can be rewritten as a number of homogeneous matrix equations (one for each degree and symmetry) for the coefficients $c_{np}$. The frequencies of the inertial modes in a triaxial ellipsoid correspond to those values of $\lambda$ or $\omega$ that allow non-zero values for $c_{np}$, i.e. that cancel the determinant of the matrix.

We now list the half-frequency of the first few inertial modes in a triaxial ellipsoid.

For the spin-over mode\footnote{reference?} (an equatorially antisymmetric mode of degree $n=2$)
\begin{equation}
\lambda^2=\frac{1-f^2}{\left(2-e^2\right) \left(2-e^2-f^2\right)}~,
\label{eq:spin-over_triaxial}
\end{equation}

The equatorially symmetric modes of degree 3 are
\tiny
\begin{equation}
\lambda^2=\frac{(1-f^2) \left(65-4 e^2-63 f^2+2 e^2f^2+12 f^4\pm\sqrt{4000-400 e^2-7800 f^2+600 e^2 f^2+5321 f^4-282 e^2 f^4-1480 f^6+e^4 f^4+40 e^2 f^6+144 f^8}\right)}{(15-4e^2-4 f^2+e^2 f^2) (15-4e^2-22 f^2+3e^2 f^2+8f^4)}
\end{equation}
\normalsize

The equatorially antisymmetric modes of degree 3 are
\begin{equation}
\lambda^2=\frac{(1-f^2) \left(7-8e^2-3 f^2+2 e^4+2e^2f^2\pm(1-e^2)\sqrt{4 (1-e^2)^2+f^4}\right)}{(3-2e^2-2 f^2+e^2f^2) (15-22e^2-4 f^2+8 e^4+3e^2 f^2)}
\end{equation}

\section{Spherical Harmonics decompositions}
\label{sec:appendixSH}
We give the spherical harmonics coefficients of the expressions that appear in Eq.~(\ref{eq:Poincare}-\ref{eq:bcPoincare}) in spherical coordinates.\footnote{Those expressions depend on the choice of normalisation. Here, we use the semi-normalised spherical harmonics for which one has
$Y_0^0=1$.}
\begin{equation}
\left(\nabla^2\phi\right)_{\ell,m}(r)=\left(\frac{d^2}{dr^2}+\frac{2}{r}\frac{d}{dr}-\frac{\ell(\ell+1)}{r^2}\right)\phi_{\ell,m}(r)
\end{equation}
\begin{align}
\left(\frac{d^2}{dz^2}\phi\right)_{\ell,m}(r)=
&\frac{\sqrt{(\ell+1)^2-m^2}\sqrt{(\ell+2)^2-m^2}}{15+16\ell+4\ell^2}\left(\frac{d^2}{dr^2}+\frac{2\ell+5}{r}\frac{d}{dr}+\frac{\ell^2+4\ell+3}{r^2}\right)\phi_{\ell+2,m}(r)\nonumber\\
&+\frac{2\ell(\ell+1)-1-2m^2}{4\ell^2+4\ell-3}\left(\frac{d^2}{dr^2}+\frac{2}{r}\frac{d}{dr}-\frac{\ell(\ell+1)}{r^2}\right)\phi_{\ell,m}(r)\nonumber\\
&+\frac{\sqrt{(\ell-1)^2-m^2}\sqrt{\ell^2-m^2}}{4\ell^2-8\ell-3}\left(\frac{d^2}{dr^2}+\frac{3-2\ell}{r}\frac{d}{dr}+\frac{\ell(\ell-2)}{r^2}\right)\phi_{\ell-2,m}(r)
\end{align}
\begin{align}
\left(({\bf \hat{r}}\cdot{\bf \hat{z}})\frac{d\phi}{dz}\right)&_{\ell,m}(r)=\nonumber\\
&\left(\frac{\sqrt{(\ell+1-m)(\ell+2-m)(\ell+m+1)(\ell+2+m)}}{4(\ell+2)^2-1}\frac{d}{dr}\right.\nonumber\\
&\left.+\frac{\sqrt{(\ell^2+5\ell+6)(\ell+3)(\ell+2-m)(\ell-m+1)(\ell+m+1)(\ell+2+m)}}{(2\ell+3)(2\ell+5)\sqrt{\ell+2}}\frac{1}{r}\right)\phi_{\ell+2,m}(r)\nonumber\\
&+\frac{1}{4\ell^2+4\ell-3}\left((-1+2\ell(\ell+1)-2m^2)\frac{d}{dr}+\frac{(\ell^2+\ell-3m^2)}{r}\right)\phi_{\ell,m}(r)\nonumber\\
&\left(\frac{\sqrt{(\ell-1-m)(\ell-m)(\ell+m-1)(\ell+m)}}{4\ell^2-8\ell-3}\frac{d}{dr}\right.\nonumber\\
&\left.+\frac{\sqrt{(\ell^2-3\ell+2)(\ell-2)(\ell-1-m)(\ell-m)(\ell+m-1)(\ell+m)}}{(4\ell^2-8\ell+3)\sqrt{\ell-1}}\frac{1}{r}\right)\phi_{\ell-2,m}(r)
\end{align}
\begin{equation}
\left(({\bf \hat{z}}\times{\bf \hat{r}})\cdot{\bf \nabla}\phi\right)_{\ell,m}(r)=\frac{2im}{r}\phi_{\ell,m}(r)
\end{equation}

\section{Oblate spheroidal coordinates}
\label{sec:spheroidal}
It is possible to extend the usage of spherical harmonics to the treatment of equations written in oblate spheroidal coordinates:
\begin{equation}
\begin{cases}
x=c~\sqrt{1+\xi^2}\sin\theta\cos\phi\\
y=c~\sqrt{1+\xi^2}\sin\theta\sin\phi\\
z=c~\xi\cos\theta~.
\end{cases}
\end{equation}
The level surfaces labelled with different constant values of $\xi$ correspond to a family of confocal spheroids with foci separated by a distance $2c > 0$ in the $xy$-plane.

\begin{figure}
\includegraphics[width=\textwidth]{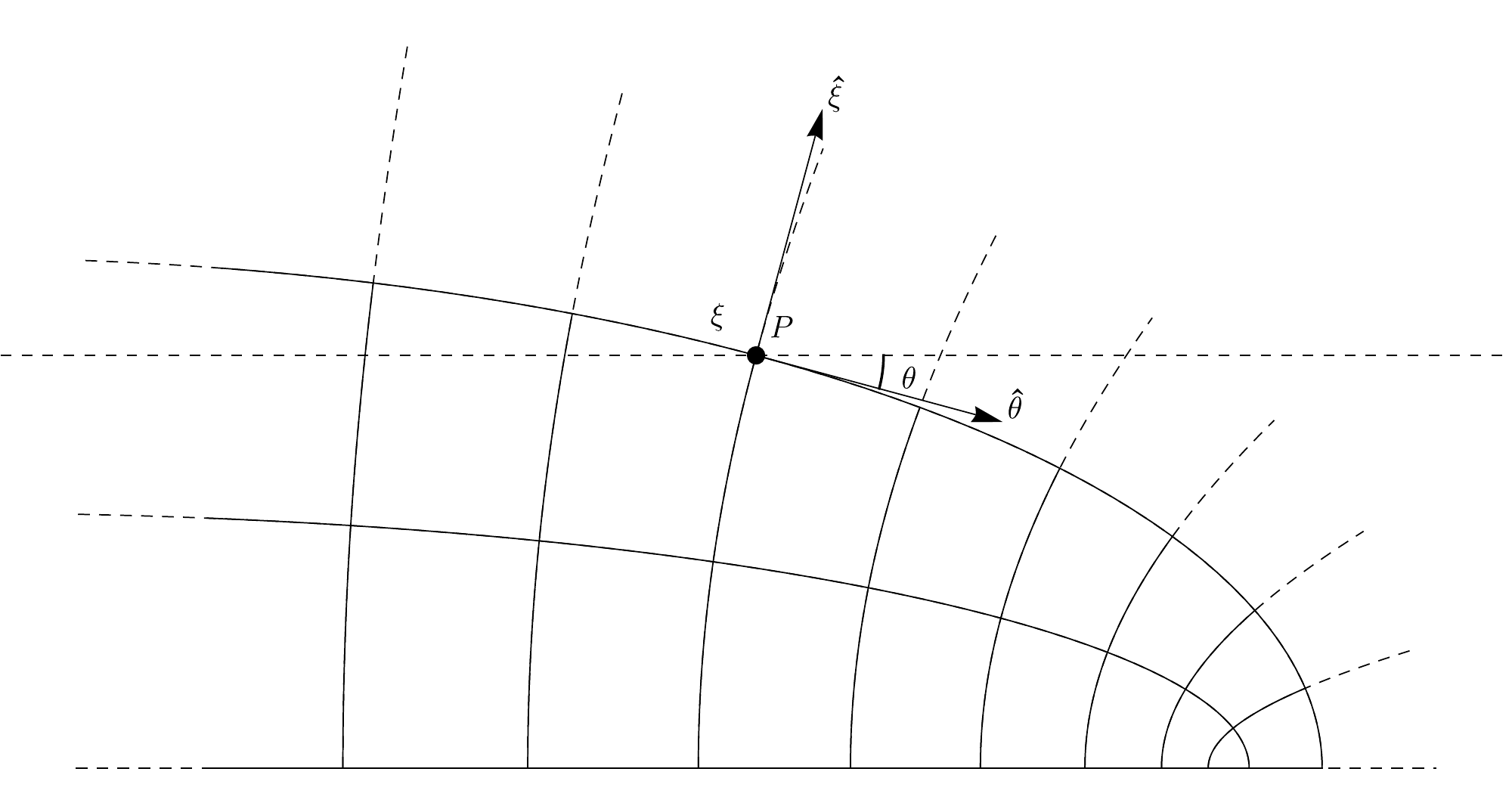}
\caption{Localisation of a point $P$ by means of its oblate spheroidal coordinates ($\xi,\theta$) at an arbitrary azimuthal angle $\phi$. level surfaces of constant $\xi$ are oblate spheroids of linear eccentricity $c>0$. The basis vector ${\bf\hat{\theta}}$ makes an angle $\theta$ with the horizontal direction.}
\label{fig:spheroidalcoordinates}
\end{figure}

A scalar function given in oblate spheroidal coordinates can be expanded on a series of spherical harmonics just as easily as if it were given in spherical coordinates. The only difference being that, the coordinate $\theta$ no longer represents the polar angle (\emph{geocentric} colatitude) but rather the angle between the basis vector ${\bf\hat{\theta}}$ and the horizontal as illustrated on Fig.~\ref{fig:spheroidalcoordinates} (\emph{geodetic} colatitude). The $\xi$ coordinate now plays a role similar to that of the radial spherical coordinate $r$. In oblate spheroidal coordinates, the (truncated) expansion of the reduced pressure reads 
\begin{equation}
p(\xi,\theta,\phi)=\sum_{\ell=0}^L\sum_{m=-\ell}^\ell p_{\ell,m}(\xi)Y_\ell^m(\theta,\phi)~.
\end{equation}
The trick to using spherical harmonics expansion effectively for the resolution of differential equations in these coordinates consists in realising that the result of any differential operator acting on a single harmonic function, $Y_\ell^m$, can be written as a finite sum of harmonic functions multiplied by some power of a \emph{common prefactor}:
\begin{equation}
\frac{1}{2(\xi^2+\cos^2\theta)}~.
\label{eq:prefactorspheroidal}
\end{equation}
When the equation to solve consists of the sum of differential operators, each with its own power of the prefactor (\ref{eq:prefactorspheroidal}), the idea is to reduce the whole equation to the same denominator. This generates products of the spherical harmonics in each numerator with some power of $\cos^2\theta$ (which has a simple harmonic expansion in terms of $Y_2^0$ and $Y_0^0$ only). These products can be reduced to a finite sum over spherical harmonics using the usual rules. This technique critically reduces the coupling between each resulting ODE which would otherwise be infinite. In practice, we use the Mathematica package \emph{TenGSHui} to expand the equations in oblate spheroidal harmonics.

\section{series expansion of an ellipsoid}
\label{sec:seriesell}

The coefficients of the expansion of the triaxial ellipsoid surface in spherical harmonics which appear in Eq.~(\ref{eq:bcSH}) are given below to second order in the squared eccentricities $e^2$ and $f^2$.
\begin{align}
\epsilon_{0,0}&=-\frac{11 e^4}{120}+\frac{e^2 f^2}{20}-\frac{e^2}{6}-\frac{11 f^4}{120}-\frac{f^2}{6}\nonumber\\
\epsilon_{2,-2}&=-\frac{1}{28} \sqrt{\frac{3}{2}} e^2 f^2+\frac{5 f^4}{28 \sqrt{6}}+\frac{f^2}{2 \sqrt{6}}\nonumber\\
\epsilon_{2,0}&=-\frac{5 e^4}{42}+\frac{e^2 f^2}{28}-\frac{e^2}{3}+\frac{5 f^4}{84}+\frac{f^2}{6}\nonumber\\
\epsilon_{2,2}&=-\frac{1}{28} \sqrt{\frac{3}{2}} e^2 f^2+\frac{5 f^4}{28 \sqrt{6}}+\frac{f^2}{2 \sqrt{6}}\nonumber\\
\epsilon_{4,-4}&=\frac{3 f^4}{8 \sqrt{70}}\nonumber\\
\epsilon_{4,-2}&=\frac{3 f^4}{28 \sqrt{10}}-\frac{3 e^2 f^2}{14 \sqrt{10}}\nonumber\\
\epsilon_{4,0}&=\frac{3 e^4}{35}-\frac{3 e^2 f^2}{35}+\frac{9 f^4}{280}\nonumber\\
\epsilon_{4,2}&=\frac{3 f^4}{28 \sqrt{10}}-\frac{3 e^2 f^2}{14 \sqrt{10}}\nonumber\\
\epsilon_{4,4}&=\frac{3 f^4}{8 \sqrt{70}}
\end{align}

\end{appendix}

\end{document}